\documentclass[12pt]{article}

\usepackage[textheight=22.5truecm,textwidth=16.8truecm,top=2.95truecm,left=2.15truecm]{geometry}
\usepackage{
amsmath,
amsthm,
amssymb,
ascmac,
bm,
tikz,
braket,
mathrsfs,
graphicx,
color,
hyperref,
cite,
titlesec,
authblk,
caption,
subfigure
}

\hypersetup{colorlinks,bookmarksopen,bookmarksnumbered,citecolor=black,linkcolor=black,linktocpage,pdfstartview=FitH,urlcolor=black}

\numberwithin{equation}{section}

{\makeatletter
\long\def\@makefntext#1{\parindent 1em\noindent 
\@hangfrom{\hbox to 1.8em{\hss$^{\@thefnmark}$}}#1}
\makeatother}
\renewcommand\thefootnote{\arabic{footnote})}

\def\fnum@figure{\textbf{\figurename\nobreakspace\thefigure}}
\def\fnum@table{\textbf{\tablename\nobreakspace\thetable}}
\long\def\@makecaption#1#2{%
  \vskip\abovecaptionskip
  \sbox\@tempboxa{\small #1. #2}%
  \ifdim \wd\@tempboxa >\hsize
    \small #1. #2\par
  \else
    \global \@minipagefalse
    \hb@xt@\hsize{\hfil\box\@tempboxa\hfil}%
  \fi
  \vskip\belowcaptionskip}
\renewcommand{\l}[0]{\left}
\renewcommand{\r}[0]{\right}
\renewcommand{\d}[0]{\mathrm{d}}

\allowdisplaybreaks

\title{\hfill\parbox{3cm}{
\normalsize
KUNS-2964
}\\[12pt]
Shooting null geodesics into holographic spacetimes
}

\author[1,2]{Shunichiro Kinoshita}
\author[1]{Keiju Murata}
\author[3]{Daichi Takeda}

\affil[1]{\it Department of Physics, College of Humanities and Sciences, Nihon University, Tokyo 156-8550, Japan}
\affil[2]{\it Department of Physics, Chuo University, Kasuga, Bunkyo-ku, Tokyo 112-8551, Japan}
\affil[3]{\it Department of Physics, Kyoto University, Kyoto 606-8502, Japan}

\date{}

\begin{document}
%Title
\maketitle

%E-mails
\renewcommand{\thefootnote}{\fnsymbol{footnote}}
\footnote[0]{
kinoshita.shunichiro@nihon-u.ac.jp,
murata.keiju@nihon-u.ac.jp,
takedai@gauge.scphys.kyoto-u.ac.jp
}
\renewcommand{\thefootnote}{\arabic{footnote}}

\begin{abstract}
We find, in the AdS/CFT, a source on the boundary which generates one wave packet drawing a null geodesic inside the bulk. Once such a wave packet dives into the bulk, it comes back to the boundary after a specific time, at which the expectation value of the corresponding boundary operator finally stands up. Since this behavior strongly reflects the existence of the holographic spacetime, our technique will be helpful in identifying holographic materials.
\end{abstract}

\newpage

\tableofcontents

\section{Introduction}\label{sec: intro}

The AdS/CFT correspondence predicts that some gravitational systems in asymptotically AdS spacetimes describe quantum phenomena in strongly coupled field theories~\cite{Maldacena:1997re,Gubser:1998bc,Witten:1998qj}.
Along with that, the application of the AdS/CFT to condensed matter systems has also been investigated in recent years \cite{Hartnoll:2009sz,Herzog:2009xv,McGreevy:2009xe,Horowitz:2010gk,Sachdev:2010ch}.
Nevertheless, no one knows if there is a real material which has its gravitational dual in our world.
The discovery of such a material will make it possible to experiment with classical or even quantum gravity in table-top experiment.
Thus, it is reasonable to propose a tool which can be used to determine whether the material has the gravitational dual.

One of the main tools proposed so far is the application of the optical imaging to materials~\cite{Hashimoto:2018okj,Hashimoto:2019jmw,Kaku:2021xqp,Zeng:2022woh,Hashimoto:2022aso,Caron-Huot:2022lff,Zeng:2023zlf}.
Let us consider a material processed to a sphere ($\mathbb S^1$ or $\mathbb S^2$) and put a local source on it.
If the material is holographic, the response to the source can be computed by the classical wave which propagates over the bulk spacetime emergent inside the sphere.
Thus if we looked into the bulk with our eyes, we would see the image of the source created by the gravitational lens.
Here, the optical imaging is a mathematical operation similar to the Fourier transformation, whose role is to provide the image that our eyes would see.
By using this, we can take a black hole image just as the Event Horizon Telescope~\cite{EventHorizonTelescope:2019dse,EventHorizonTelescope:2022} did, or can catch a signal of the emergence of the pure AdS geometry.

From the viewpoint of the eikonal approximation, the above idea came from the question as to how null-geodesic congruences going from the boundary to boundary can be seen holographically.
In this paper, we rather focus on making each null geodesic, and retrieving it on the boundary. (See \cite{Kaku:2022hcg} for the creation of the timelike geodesic.)
We prepare a source parameterized by its frequency and wavenumbers, and see that it generates a wave packet going along a null-geodesic orbit.
As shown in \cite{Berenstein:2019tcs,Berenstein:2020vlp}, localized states in the AdS bulk can be realized by applying nonlocal operators to states in the dual quantum field theory (QFT). 
We will provides an explicit way to create such sates by using external sources in QFT as has been done in~\cite{Kaku:2022hcg}.

Since the technique allows us to shoot null geodesics at will, we are to obtain another way of confirming spacetime emergence.
Once the source is turned on, a wave packet propagates inside the bulk and it will not come back to the boundary for a while.
On the boundary, the expectation value of the corresponding operator will stay quiet during that time.
Then it will suddenly stand up when the geodesic reaches the boundary.
The phenomenon happening on the boundary is so unique that it can be a signal in searching holographic materials.
Besides, all we have to do is just to measure the time lapse $\Delta t$ from the source is turned on until the response stands up.
For example, if the bulk is the pure AdS, any null geodesic reaches the boundary with $\Delta t = \pi L$, where $L$ is the AdS radius, or if a black hole exists, $\Delta t$ experiences a rapid increase according to the change of a wavenumber of the source.
We will show this more in detail later.

The organization of this paper is as follows.
We first study the characteristic of $\Delta t$ in section \ref{sec: time lapse} for the pure AdS$_{3,4}$ and Schwarzschild-AdS$_4$.
Next, in section \ref{sec: shooting}, we introduce the source, which we show generates a wave packet along a null geodesic.
Here we will also check the above expectant behavior of the response function.
Section \ref{sec: summary} is devoted for summary and discussions.
In appendix \ref{numerical}, the details of our numerical computation is explained.
In appendix \ref{app: AdS}, for the pure AdS$_3$, we analytically solve the equation of motion appearing in section \ref{sec: shooting}.

\section{Null geodesics in asymptotically AdS geometries}\label{sec: time lapse}

\begin{figure}
  \centering
\subfigure[Effective potential]
 {\includegraphics[scale=0.55]{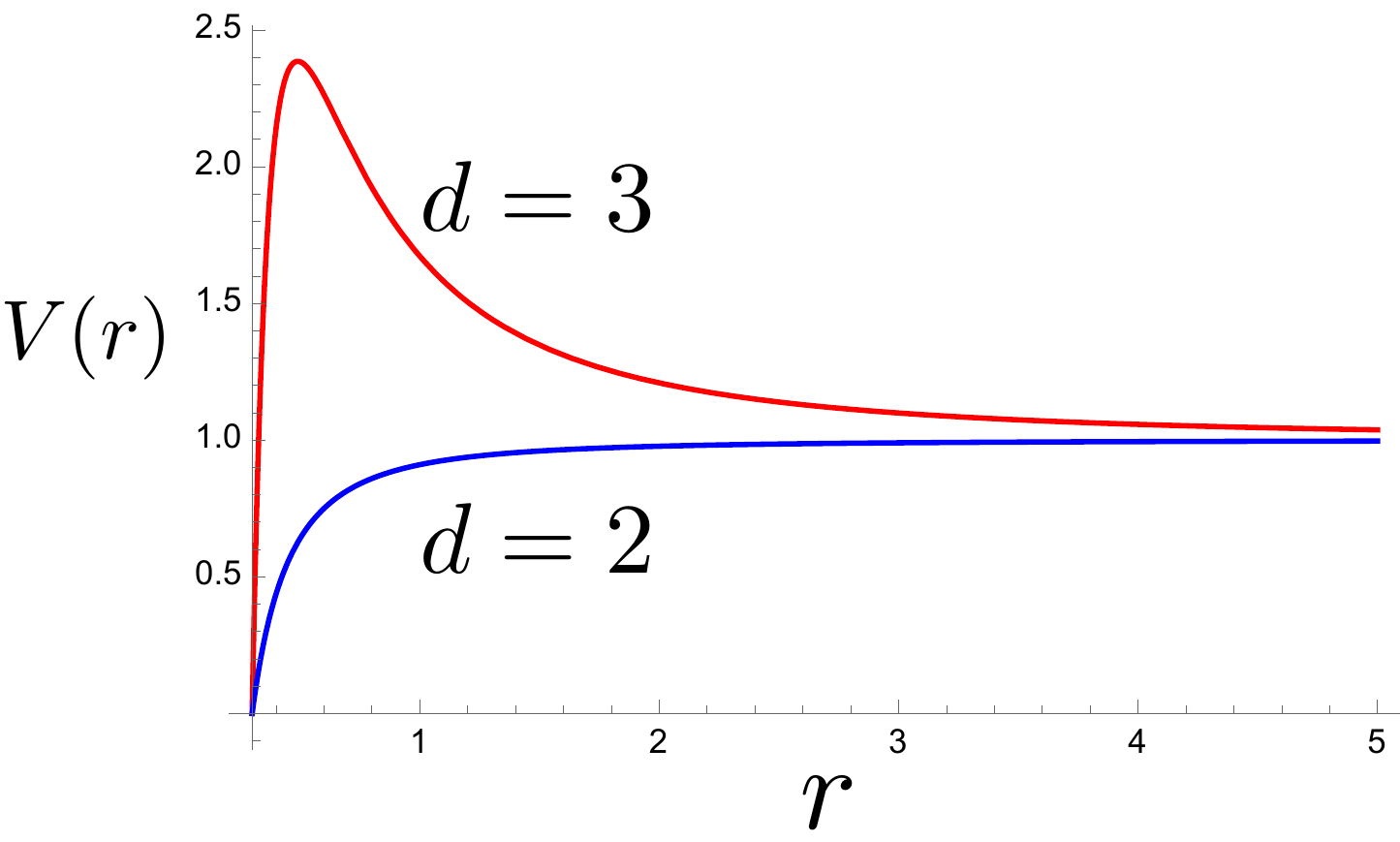}\label{Veff}
  }
  \subfigure[Null orbit]
 {\includegraphics[scale=0.4]{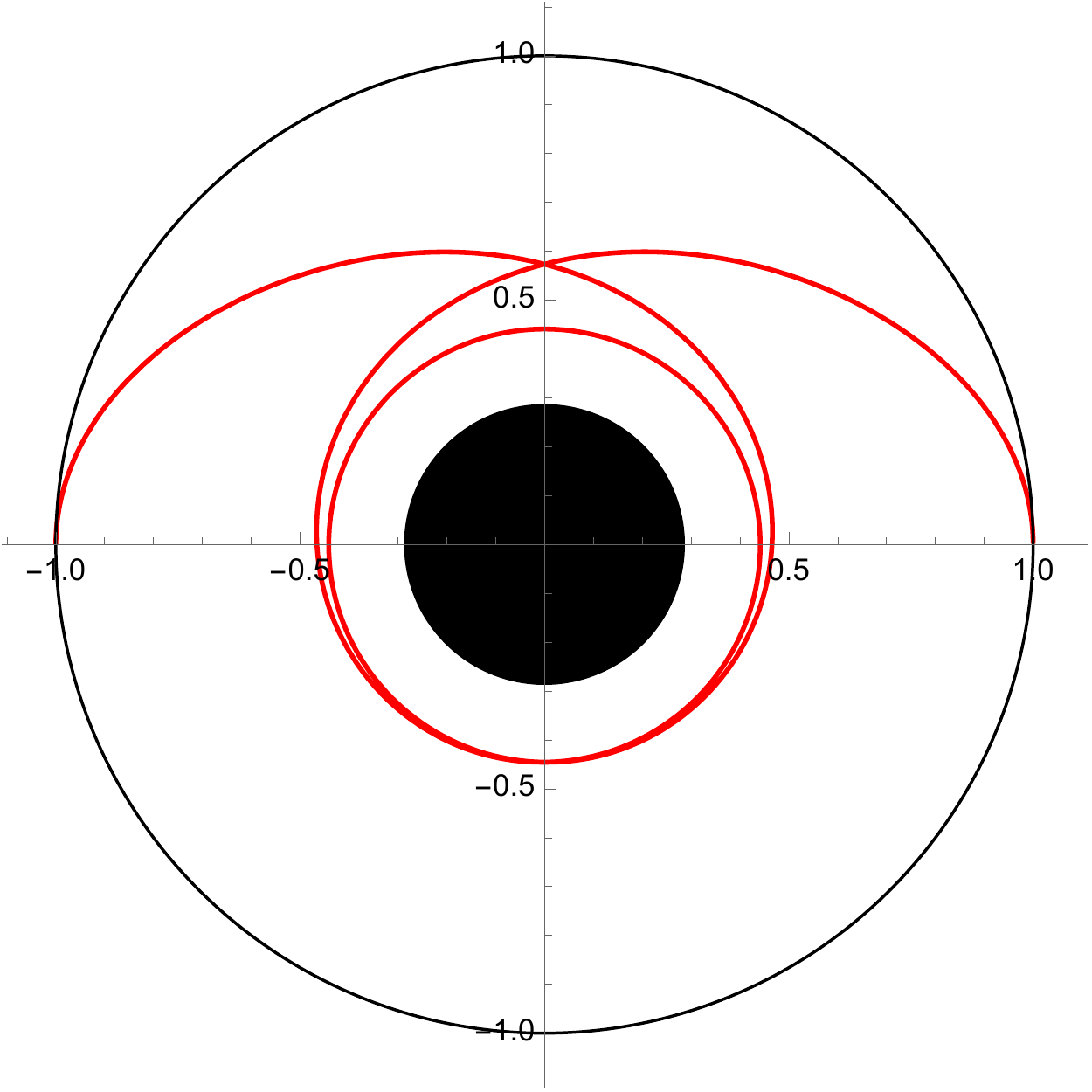}\label{nullorfbit}
  }
 \caption{
(a) Typical profiles of the effective potential for the BTZ ($d=2$) and Sch-AdS$_4$ ($d=3$) with $r_h=0.3$.
(b) Orbit of the null geodesic in the Sch-AdS$_4$ with $L=1$ and $r_h=0.3$. The specific angular momentum is $m=0.647459$. The horizontal and vertical axes are $x$ and $y$ defined in Eq.~(\ref{xy}).}
\end{figure}

Once the null geodesic is created in the AdS bulk spacetime, we can probe the bulk geometry and extract some information about the bulk metric.
For example, when there is a black hole in the bulk, the null geodesic is strongly bent and goes around the black hole (see Fig.~\ref{nullorfbit}).
If the parameter of the null geodesic is fine-tuned, it can circle around the black hole infinitely. 
The surface on which the null geodesic can orbit for infinite times is called the photon surface.
We demonstrate that the evidence of the existence of the  photon surface can be obtained by observing the time lapse $\Delta t$ between the injection and arrival of the null geodesic at the AdS boundary.

We consider the Ba\~{n}ados-Teitelboim-Zanelli (BTZ) and Schwarzschild-AdS$_4$ (Sch-AdS$_4$) spacetimes:
\begin{align}
	\d s^2 = -f(r) \d t^2 + f(r)^{-1}\d r^2 + r^2\d \Omega_{d-1}^2,
	\qquad
	\d \Omega_1^2 = \d\phi^2,\qquad
    \d \Omega_2^2 = \d\theta^2 + \sin^2\theta \d\phi^2.
    \label{eq: metric}
\end{align}
Here $d\Omega^2_{d-1}$ is the standard metric of the unit sphere $\mathbb S^{d-1}$, and we consider $d=2$ or $3$. The function $f(r)$ is given by
\begin{equation}
 f(r)=\frac{r^2-r_h^2}{L^2} \quad (d=2)\ ,\quad 
f(r)=1+\frac{r^2}{L^2}-\frac{r_h(1+r_h^2/L^2)}{r}\quad (d=3)\ ,
\end{equation}
where $r=r_h$ is the locus of the event horizon and $L$ is the AdS radius. 
%We take the unit of $L=1$ in our actual calculations.
%, since we are now focusing on real materials.
Using the spherical symmetry, we can put any geodesic on the equatorial plane, $\theta=\pi/2$, without loss of generality.
In terms of $t$ and $\phi$, time-translation symmetry and axisymmetry of the spacetimes yield two conserved quantities along null geodesics.
%From conservation laws of the energy the angular momentum and null condition, for a null geodesic are then given as
Then, for a null geodesic, these conservation laws and the null condition provide 
\begin{align}
	f(r) \dot t = \Omega,\qquad
	r^2 \dot \phi = M,\qquad
	0 = -f(r) \dot t^2 + f(r)^{-1}\dot r^2 + r^2 \dot \phi^2,\label{eq: geodesic equations}
\end{align}
where the dot denotes the derivative by an affine parameter $\lambda$. Conserved quantities $\Omega$ and $M$ correspond to the energy and angular momentum of the null geodesic, respectively. Eliminating $\dot{t}$ and $\dot{\phi}$ from the above equations and %applying the affine transformation 
rescaling the affine parameter as 
$\lambda\to \lambda/M$, we obtain 
%and we have fixed the normalization of $\lambda$ by setting the r.h.s.\ of the first equation $1$. The, $m$ represents the angular momentum per unit energy.
\begin{equation}
 \dot{r}^2+V(r)=\frac{1}{m^2}\ ,\quad V(r)=\frac{f(r)}{r^2}\ ,
\end{equation}
where we have introduced the specific angular momentum (i.e., the angular momentum per unit energy) as
\begin{equation}
 m=\frac{M}{\Omega}\ . \label{eq:specificAM}
\end{equation}
Typical profiles of the effective potential is shown in Fig.~\ref{Veff}. For BTZ ($d=2$), the effective potential has no local maximum and any null geodesic falls into the black hole. For Sch-AdS$_4$ ($d=3$), the effective potential has the maximum value $V_\textrm{max}$. For $1/L^2<1/m^2<V_\textrm{max}$, the null geodesic injected from the AdS boundary bounces back by the potential barrier and returns again to the AdS boundary. For $1/m^2=V_\textrm{max}$, the unstable circular orbit on the photon sphere is realized. Solving Eq.~(\ref{eq: geodesic equations}) for a given $m$, we obtain an orbit of the null geodesic. Figure~\ref{nullorfbit} shows a null orbit when we tune the value of $m$ so that $1/m^2$ is close to (but smaller than) $V_\textrm{max}$. 
In Fig.~\ref{nullorfbit}, we have introduced ``Cartesian'' coordinates of the horizontal and vertical axes $(x,y)$ as 
\begin{equation}
  x=\frac{r}{\sqrt{r^2+L^2}} \cos\phi\ ,\quad y=\frac{r}{\sqrt{r^2+L^2}} \sin\phi\ ,
\label{xy}
\end{equation}
in order to compactify the AdS space, 
%We introduced ``Cartesian'' coordinates $(x,y)$ as 
%\begin{equation}
%  x=\frac{r}{\sqrt{r^2+L^2}} \cos\phi\ ,\quad y=\frac{r}{\sqrt{r^2+L^2}} \sin\phi\ .
%\label{xy}
%\end{equation}
%The horizontal and vertical axes in Fig.\ref{nullorfbit} is $x$ and $y$, respectively. 
%In the $(x,y)$-coordinates, 
where the AdS boundary is located at $x^2+y^2=1$.

\begin{figure}[t]
	\centering
	\begin{minipage}{0.4\columnwidth}
		\includegraphics[width = 6.5cm]{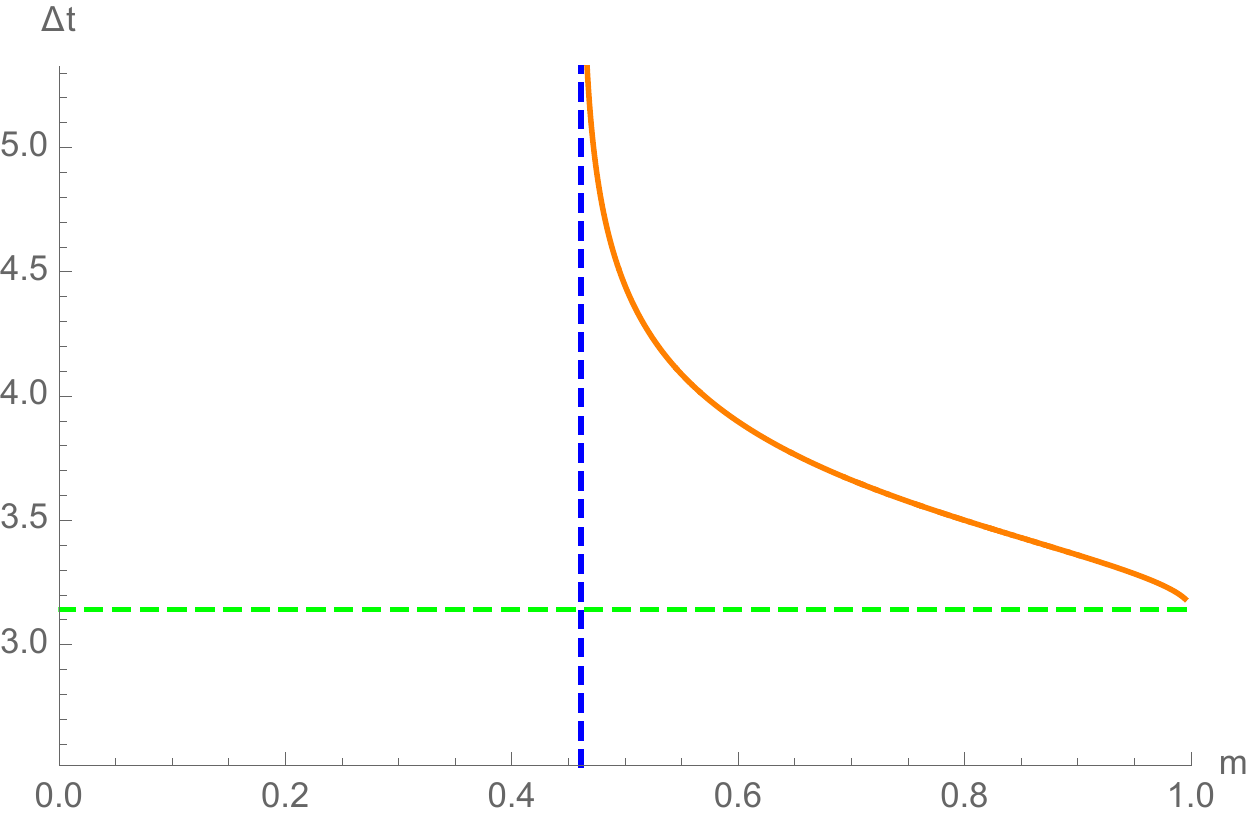}
	\end{minipage}
	\hspace{36pt}
	\begin{minipage}{0.4\columnwidth}
		\includegraphics[width = 6.5cm]{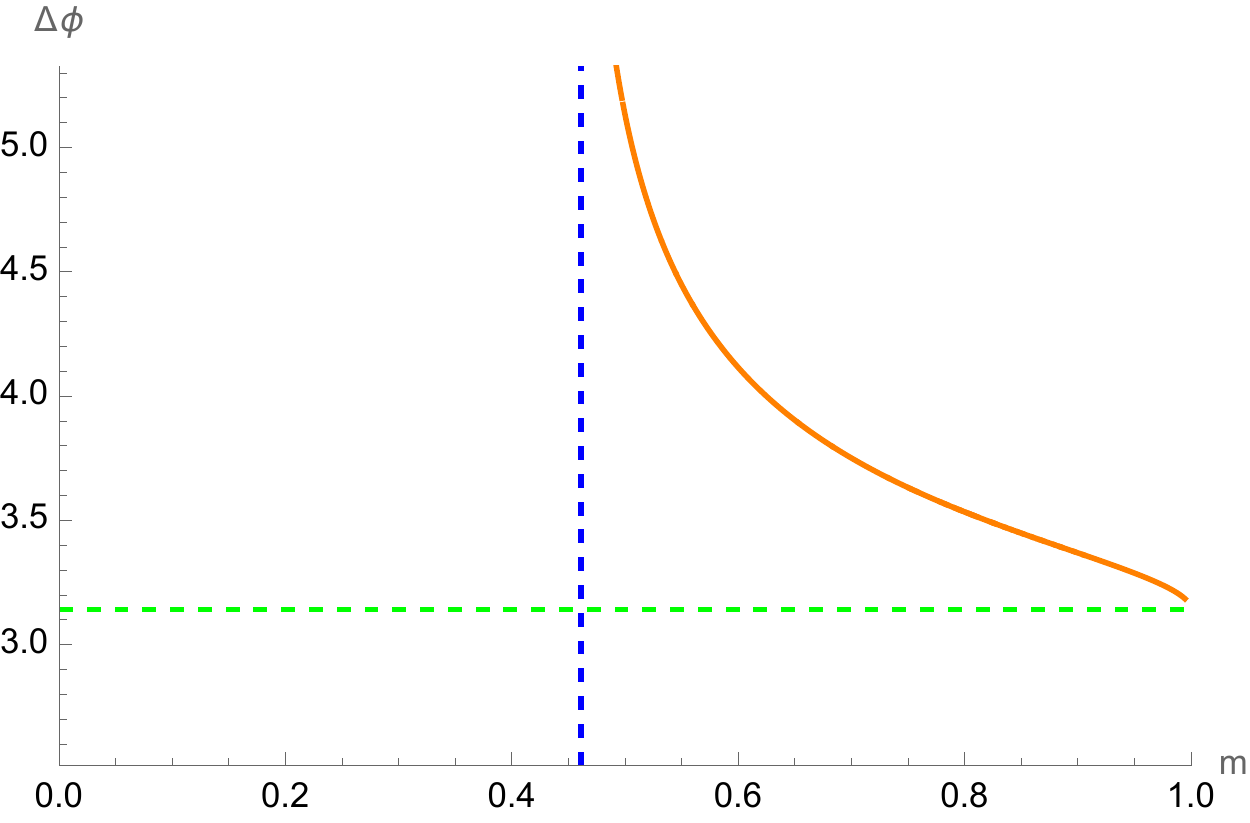}
	\end{minipage}
	\caption{The plots of $\Delta t$ and $\Delta \phi$ for Schwarzschild-AdS$_4$.
	The parameters are set as $L=1$, $r_h=0.193$. %\mu =0.1$.
	The solid lines (orange) are the $m$-dependence of $\Delta t$ and $\Delta \phi$, the horizontal dashed lines (green) are those for the pure AdS, $\pi$, and the vertical dashed lines (blue) are the threshold of $m$, below which the geodesic falls into the black hole.
	When $\Delta \phi$ is larger than $2\pi$, that means the geodesic has turned around the black hole more than once.}
	\label{fig: plot of delta}
\end{figure}

%Eliminating $\lambda$, we use $r$ for our purpose.
Equations~\eqref{eq: geodesic equations} can be rewritten into
\begin{align}
	\frac{\d t}{\d r} = \frac{r}{f(r)\sqrt{r^2-m^2f(r)}},\qquad
	\frac{\d \phi}{\d r} = \frac{m}{r\sqrt{r^2-m^2f(r)}}.
\end{align}
Let $r_\mathrm{min}$ be the maximum root of $r^2-m^2f(r)$, that is, $V(r_\mathrm{min})=1/m^2$.
Then, the time and the angle of the geodesic coming back to the boundary are 
\begin{align}
	\Delta t = 2\int_{r_\mathrm{min}}^{\infty}\frac{r\,\d r}{f(r)\sqrt{r^2-m^2f(r)}},\qquad
	\Delta \phi= 2\int_{r_\mathrm{min}}^{\infty}\frac{m\,\d r}{r\sqrt{r^2-m^2f(r)}}.\label{eq: delta}
\end{align}
For the pure AdS, the metric is given by $f(r) = (r^2+L^2)/L^2$.
In this case, regardless of the dimension and $m\in (-L,L)$, we have $\Delta t = L\Delta \phi = \pi L$.
For the Schwarzschild-AdS$_4$, the integrals \eqref{eq: delta} can numerically be computed.
Figure~\ref{fig: plot of delta} shows $\Delta t$ and $\Delta \phi$ as functions of $m$ for $L=1$ and $r_h=0.193$. 
As $m$ approaches a critical value $m_\ast\equiv 1/\sqrt{V_\textrm{max}}$, %$$m\to m_\ast$, 
$\Delta t$ and $\Delta \phi$ diverge. This divergence originates from the existence of the photon surface: Since the null geodesic wanders around the photon surface, it takes long time to arrive at the boundary.
%If we can observe such time delay, it would be a strong evidence of the existence of the photon surface in the bulk.

In section \ref{sec: shooting}, we will study a source that generates one null geodesic.
As was mentioned in section \ref{sec: intro}, the response $\braket{O}_J$ to the source $J$ stands when the geodesic arrives at the boundary.
The source $J$ contains the parameter $m$ (see section \ref{sec: shooting}), and Fig.~\ref{fig: plot of delta} shows when and where we get the pulse of $\braket{O}_J$ with $m$ fixed.
In our strategy, we do not need to process the data of $\braket{O}_J$, or even to care about the value of $\braket{O}_J$ itself.
All we need is the behavior of $(\Delta t,\Delta \phi)$.
If we always find $(\Delta t,\Delta \phi)$ constant, in particular $\Delta \phi=\pi$, then that is a strong evidence of the material having the pure AdS as its dual spacetime.
If $(\Delta t,\Delta \phi)$ grows up as $m$ approaches a certain value $m_\ast$, below which no pulse is detected, then the material will be dual to a black hole spacetime.
Those behaviors cannot be expected without knowing the emergent spacetime, because $m$ is, in $J$, nothing but the ratio of the wavenumber to the frequency, as we will see below.

\section{The source to generate a null geodesic}\label{sec: shooting}
We have already seen that, if a material is dual to an asymptotically AdS spacetime, null geodesics in the bulk let a boundary operator behave peculiarly.
To use this in searching holographic materials, we have to develop a way to create null geodesics in the bulk by operating the boundary.
Here we use, as an example, a massless scalar field and demonstrate it.
The same method works for other fields with mass or spin, as long as the frequency of the source is kept sufficiently larger than the mass (because if so, the eikonal approximation is valid).

As a model of the bulk theory, let us consider the following scalar theory on an asymptotically AdS spacetime:
\begin{align}
	I = -\frac{1}{2}\int \d^{d+1}x\sqrt{-g}\,g^{\mu\nu }\nabla_\mu \Phi \nabla_\nu\Phi,\label{eq: action}
\end{align}
where $\nabla_\mu$ denotes the covariant derivative associated with the metric $g_{\mu\nu}$.
Near the AdS boundary, the scalar field behaves as
\begin{equation}
 \Phi(t,r,\Theta_{d-1}) = J(t,\Theta_{d-1}) + \frac{\langle \mathcal{O}(t,\Theta_{d-1})\rangle_J}{r^d}+\cdots\ , 
\end{equation}
where $\Theta_{d-1}$ denote standard spherical coordinates on $\mathbb{S}^{d-1}$ such as $\Theta_1=\phi$ and $\Theta_2=(\theta,\phi)$ in (\ref{eq: metric}). In the AdS/CFT, $J$ corresponds to the source coupling to the scalar operator $\mathcal{O}$~\cite{Klebanov:1999tb}. The response to the source $J$ appears at the sub-leading term in the asymptotic expansion of the bulk field.
In the gravity side, the function $J$ is just the boundary condition at the infinity. We assume that the scalar field is initially trivial, $\Phi|_{t=-\infty}=0$, and create the null geodesic choosing the functional profile of $J$ appropriately.

To understand the relation between fields and null geodesics, 
let us consider the eikonal approximation of the scalar field.
We put 
\begin{equation}
 \Phi = a(x)e^{iS(x)}\ ,
\label{eik}
\end{equation}
and assume that the phase $S(x)$ is a highly oscillatory function, i.e, $\nabla_\mu S(x)$ is sufficiently large. 
Then the equation of motion for the scalar field, $\nabla_\mu \nabla^\mu \Phi=0$, is reduced to $\nabla_\mu S \nabla^\mu S = 0$ in the leading order. Introducing the ($d+1$)-momentum as 
\begin{equation}
 k_\mu = \nabla_\mu S\ ,
\label{kdelS}
\end{equation}
we have the null condition $k_\mu k^\mu=0$. 
In addition, applying $\nabla_\nu$ to this and using $\nabla_\nu \nabla_\mu S = \nabla_\mu\nabla_\nu S$, we also obtain the geodesic equation $k^\nu \nabla_\nu k_\mu=0$. From Eq.~(\ref{kdelS}), we find that the null geodesic with energy $-k_t=\Omega$ and angular momentum $k_\phi=M$ corresponds to the scalar field whose phase is given by 
\begin{equation}
 S(x)\sim -\Omega t + M \phi\ .
\label{Sform}
\end{equation}
Moreover, since the spacetime admits the time-translational and axial Killing vectors, $(\partial/\partial t)^\mu$ and $(\partial/\partial\phi)^\mu$, such a scalar field with the above phase form $\Phi \sim e^{iS(x)}$ becomes a single eigenmode, satisfying 
$\mathcal{L}_{\partial_t}\Phi = -i\Omega \Phi$ and 
$\mathcal{L}_{\partial_\phi}\Phi = iM\Phi$.
Here $\mathcal{L}_\xi$ is the Lie derivative with respect to a Killing vector $\xi$.
%, and hence we regard the normal to the phase-constant surface,  $u_\mu = \partial_\mu S$, as the velocity.
Note that the larger $\Omega$ is, the more valid the eikonal approximation is, in general.

On the basis of the above analysis in the eikonal approximation, 
we propose a source on the AdS boundary to create a wave packet along the null geodesic in the asymptotically AdS spacetime as 
\begin{align}
	&d=2: J(t,\phi) = \frac{1}{2\pi \sigma_t \sigma_\phi}\exp\left[-i\Omega t + i M \phi
	-\frac{t^2}{2\sigma_t^2}-\frac{L^2\phi^2}{2\sigma_\phi^2}
	 \right]\ ,\label{eq: the source}\\
	&d=3: J(t,\theta,\phi) = \frac{1}{(2\pi)^{3/2} \sigma_t \sigma_\theta \sigma_\phi}\exp\left[-i\Omega t + i M \phi
	-\frac{t^2}{2\sigma_t^2}-\frac{L^2(\theta-\pi/2)^2}{2\sigma_\theta^2}-\frac{L^2\phi^2}{2\sigma_\phi^2}
	 \right]\ .\label{eq: the source4d}
\end{align}
When the spacetime is spherically symmetric, geodesics of our interest are those moving on the equatorial plane $\theta = \pi/2$. 
Thus, we have considered the momentum only along $\phi$-direction.

The above source function typically has the frequency $\Omega$ and wavenumber $M$ along $\phi$-direction. Thus, we can expect that the bulk field generated by the above boundary condition has the phase as in Eq.~(\ref{Sform}) and almost becomes a single eigenmode with the frequency $\Omega$ and the angular momentum $M$. Furthermore, to obtain a localized configuration of the scalar field, the amplitude of the source should be localized along time and angular directions. We need conditions $\sigma_t, \sigma_\theta,\sigma_\phi \ll \ell_\textrm{curv}$ to sufficiently localize the scalar field in the bulk, where $\ell_\textrm{curv}$ is the curvature scale of the bulk spacetime. This is typically given by $\ell_\textrm{curv}\sim L$ in our setup. Meanwhile, in the frequency domain, the source function has the width $\sim 1/\sigma_t$ around $\Omega$. The bulk field should be almost a single mode with the frequency $\sim\Omega$ and the angular momentum $\sim M$. The condition that the scalar field is localized in the momentum space is given by $1/\sigma_t \ll\Omega$. Similarly, we also have $1/\sigma_\theta, 1/\sigma_\phi \ll \Omega = M/m$, where $m$ is the specific angular momentum as in (\ref{eq:specificAM}). 
Therefore, the conditions for parameters in the source are summarized as 
\begin{equation}
 \frac{1}{\ell_\textrm{curv}} \ll \frac{1}{\sigma_t}\ ,\frac{1}{\sigma_\theta}\ ,\frac{1}{\sigma_\phi}\ll \Omega \ .
\end{equation}
For a given $m$, if we take a sufficiently large $\Omega$ under the above conditions, we can obtain a wave packet that is localized in both the real and momentum spaces.
Because a typical deviation of the angular momentum is given by $\Delta M \sim L/\sigma_\phi$, 
a deviation of the specific angular momentum $m$ becomes $\Delta m \sim L/\sigma_\phi \Omega$.
Thus, the above condition $\sigma_\phi\Omega \gg 1$ implies that the deviation of the specific angular momentum is so small.

\begin{figure}[p]
    \centering
	AdS$_3$ ($m=0.4$, $M = 40$, $\Omega = 100$, $\sigma_t = \sigma_\phi = 0.1$)\\[12pt]
	\begin{minipage}{0.2\columnwidth}
         \centering
	    \includegraphics[width = 3.5cm]{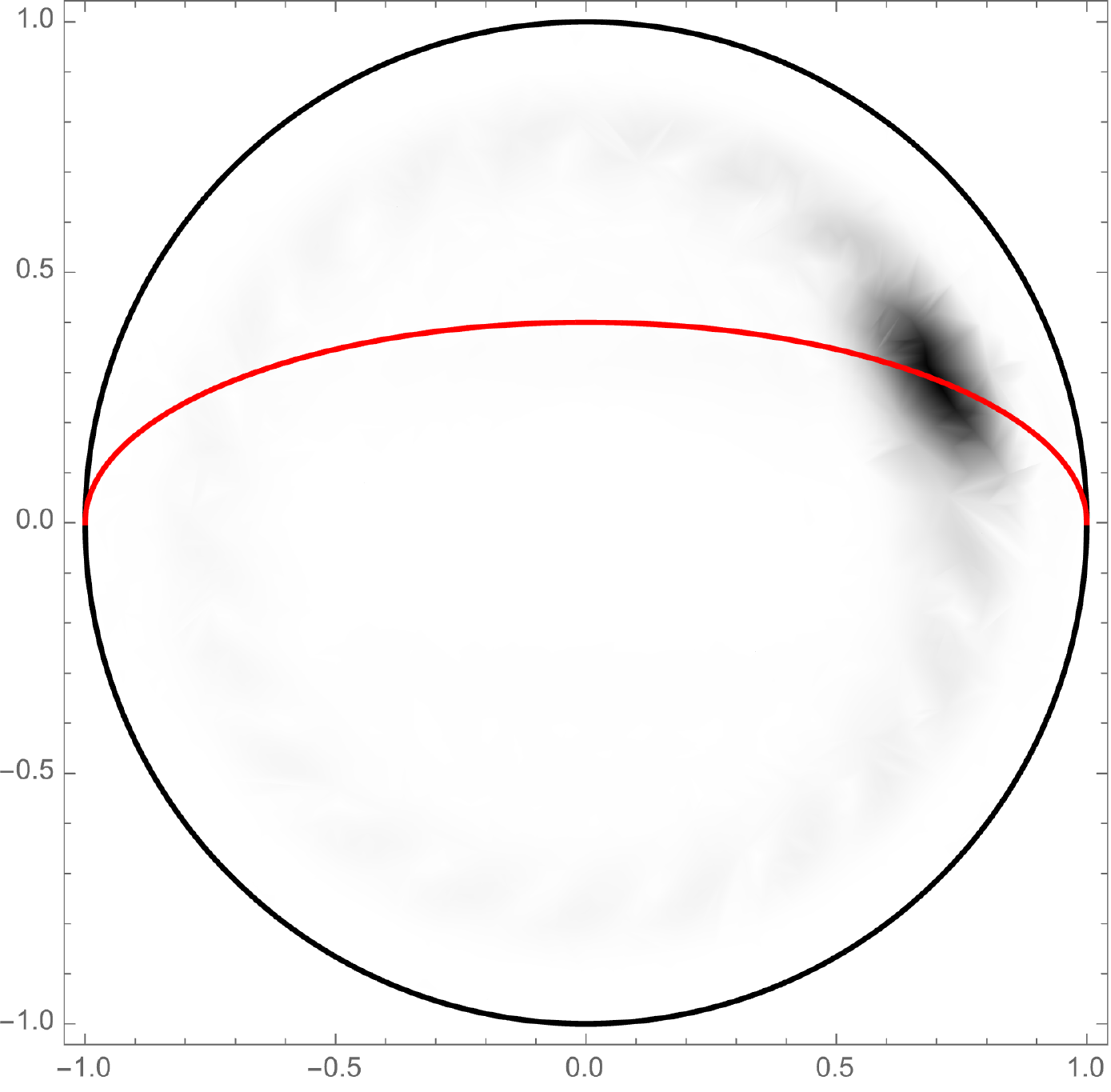}\\
        $t=\pi/4$
	\end{minipage}
     \hspace{6pt}
 	\begin{minipage}{0.2\columnwidth}
         \centering
	    \includegraphics[width = 3.5cm]{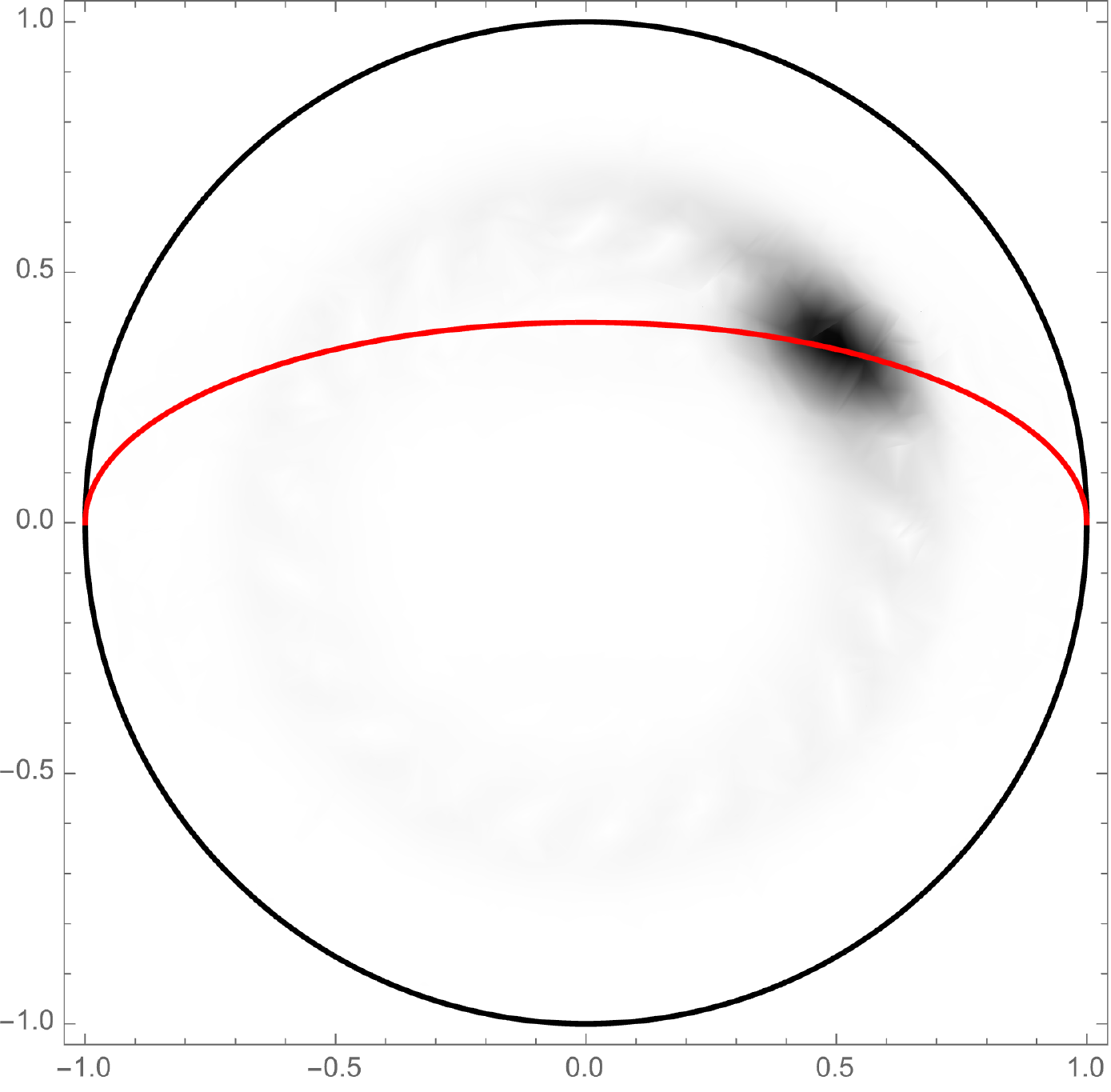}\\
        $t=\pi/3$
	\end{minipage}
     \hspace{6pt}
 	\begin{minipage}{0.2\columnwidth}
         \centering
	    \includegraphics[width = 3.5cm]{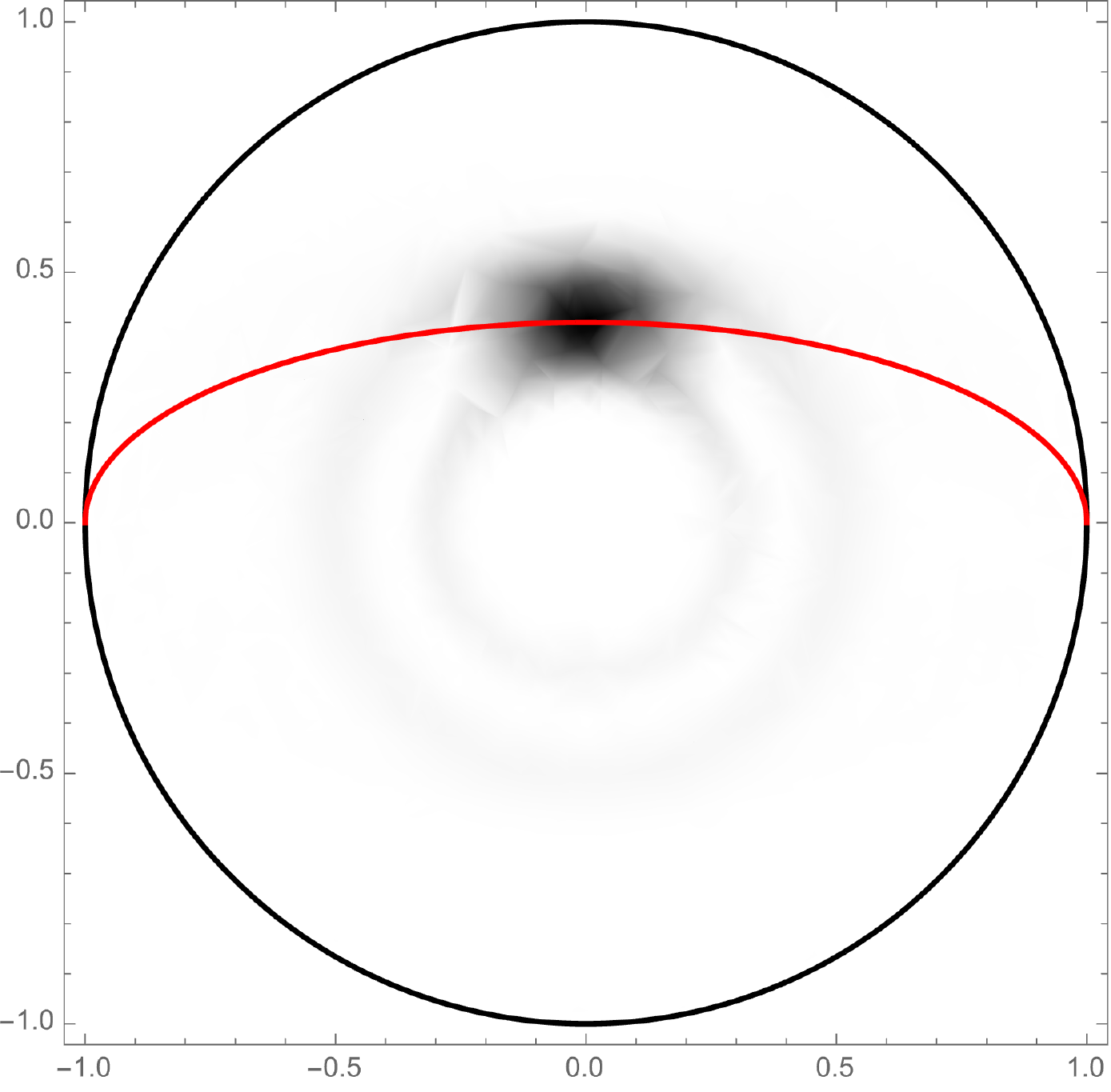}\\
        $t=\pi/2$
	\end{minipage}
    \hspace{6pt}
 	\begin{minipage}{0.2\columnwidth}
         \centering
	    \includegraphics[width = 3.5cm]{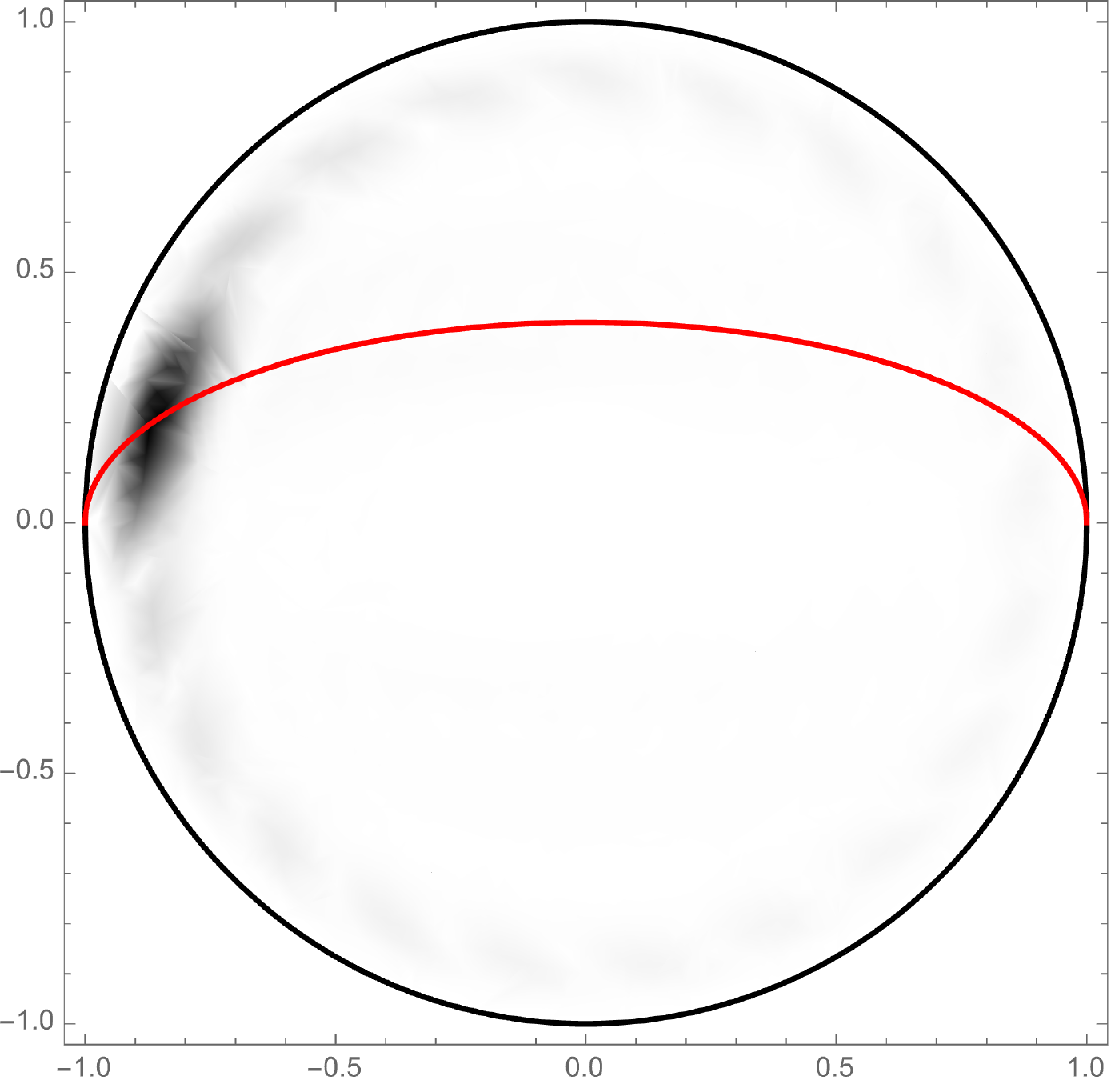}\\
        $t=5\pi/6$
	\end{minipage}\\[24pt]
 
	BTZ ($r_h=0.3$, $m=0.6$, $M = 30$, $\Omega = 50$, $\sigma_t = 0.3$, $\sigma_\phi = 0.1$)\\[12pt]
	\begin{minipage}{0.2\columnwidth}
         \centering
	    \includegraphics[width = 3.5cm]{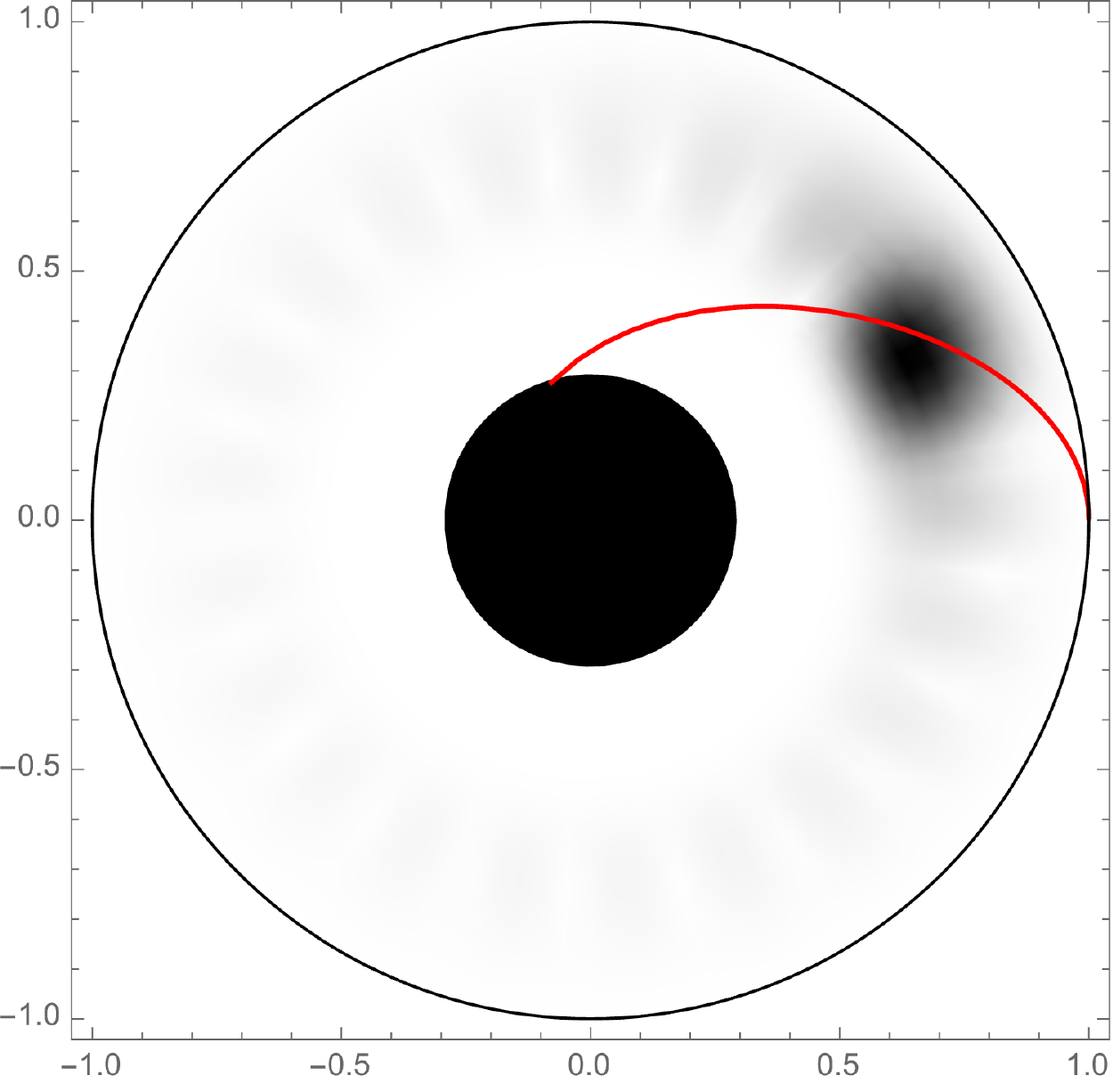}\\
        $t=0.90$
	\end{minipage}
     \hspace{6pt}
 	\begin{minipage}{0.2\columnwidth}
         \centering
	    \includegraphics[width = 3.5cm]{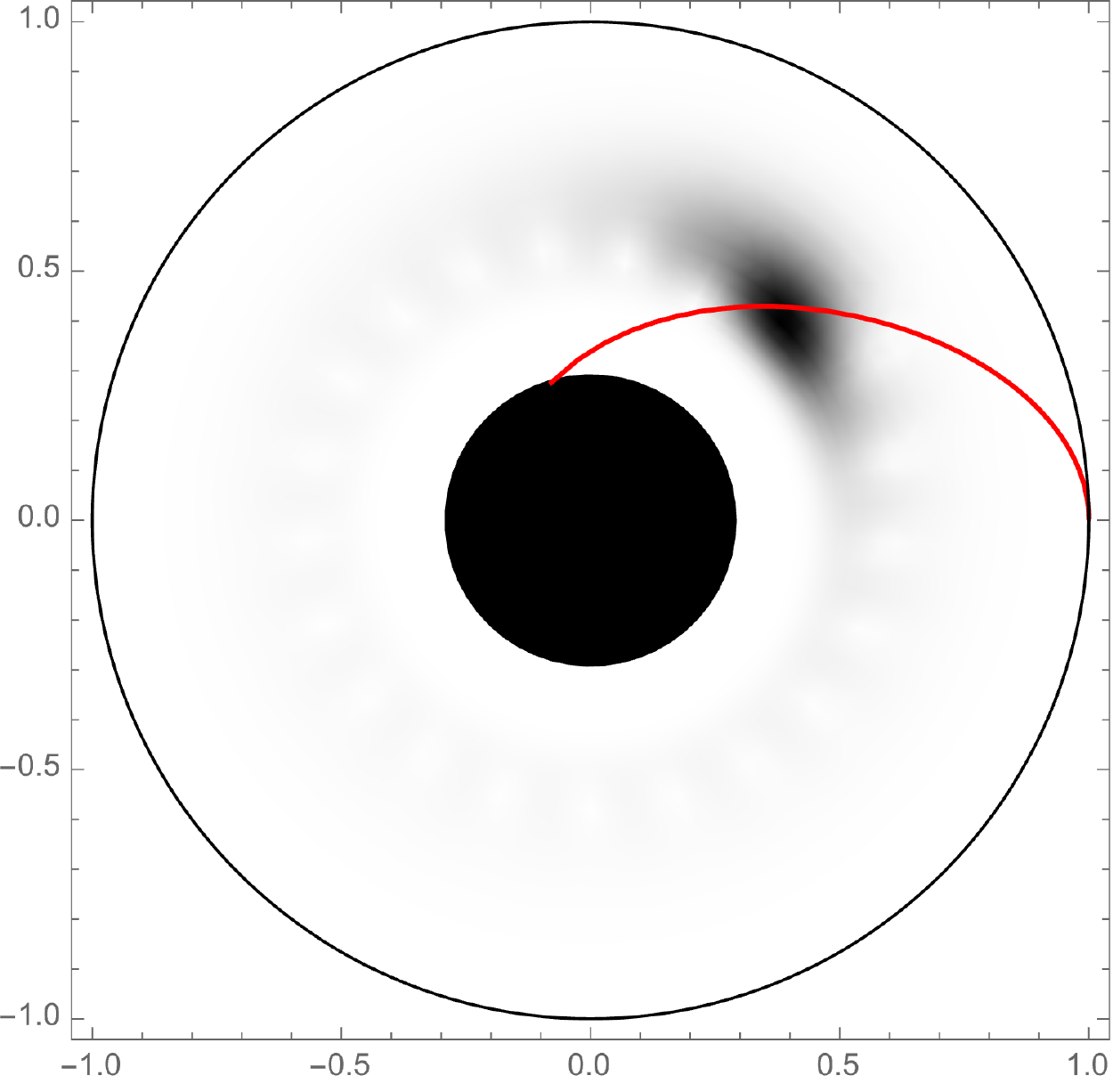}\\
         $t=1.65$
	\end{minipage}
     \hspace{6pt}
 	\begin{minipage}{0.2\columnwidth}
         \centering
	    \includegraphics[width = 3.5cm]{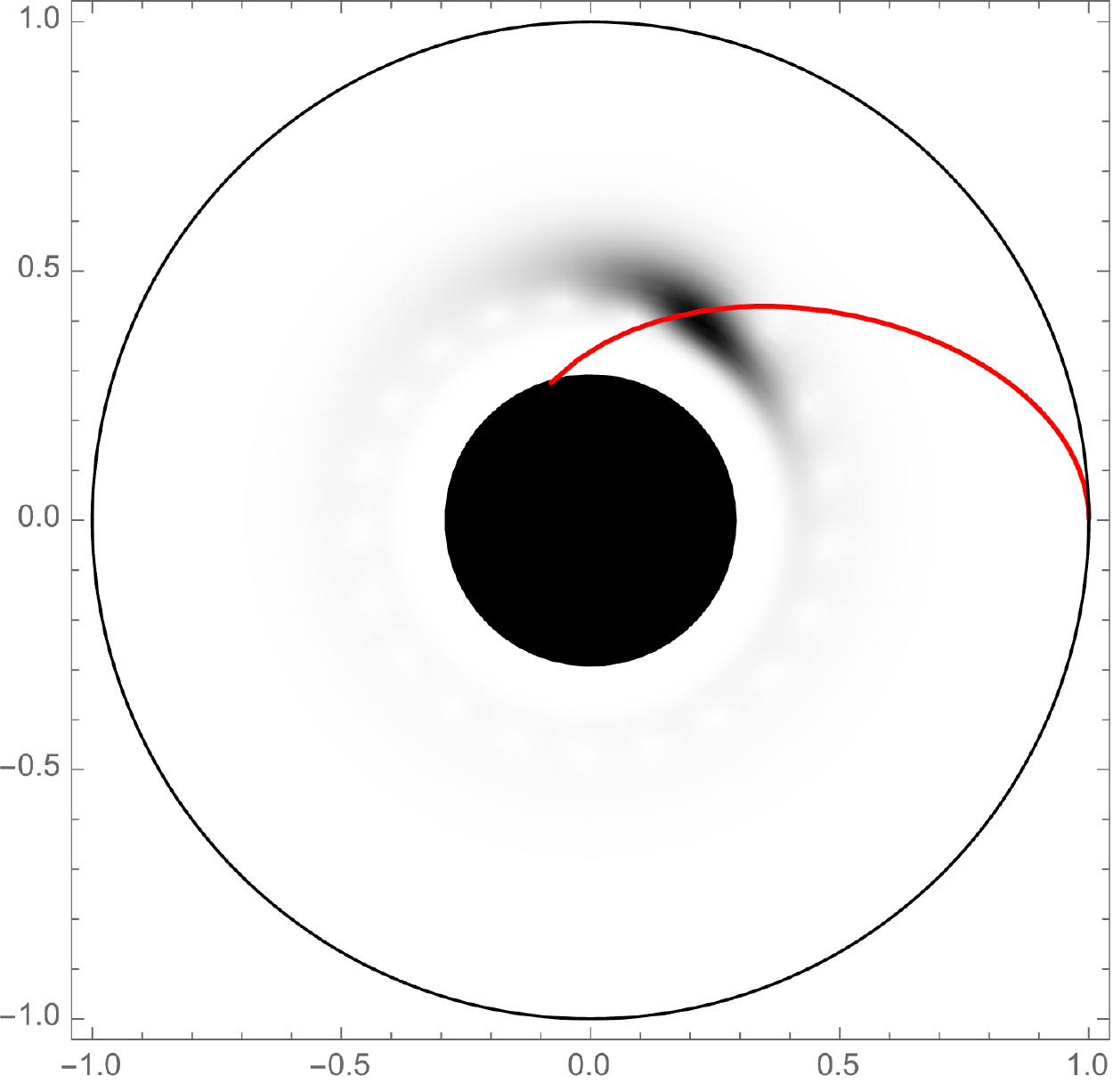}\\
        $t=2.40$
	\end{minipage}
    \hspace{6pt}
 	\begin{minipage}{0.2\columnwidth}
         \centering
	    \includegraphics[width = 3.5cm]{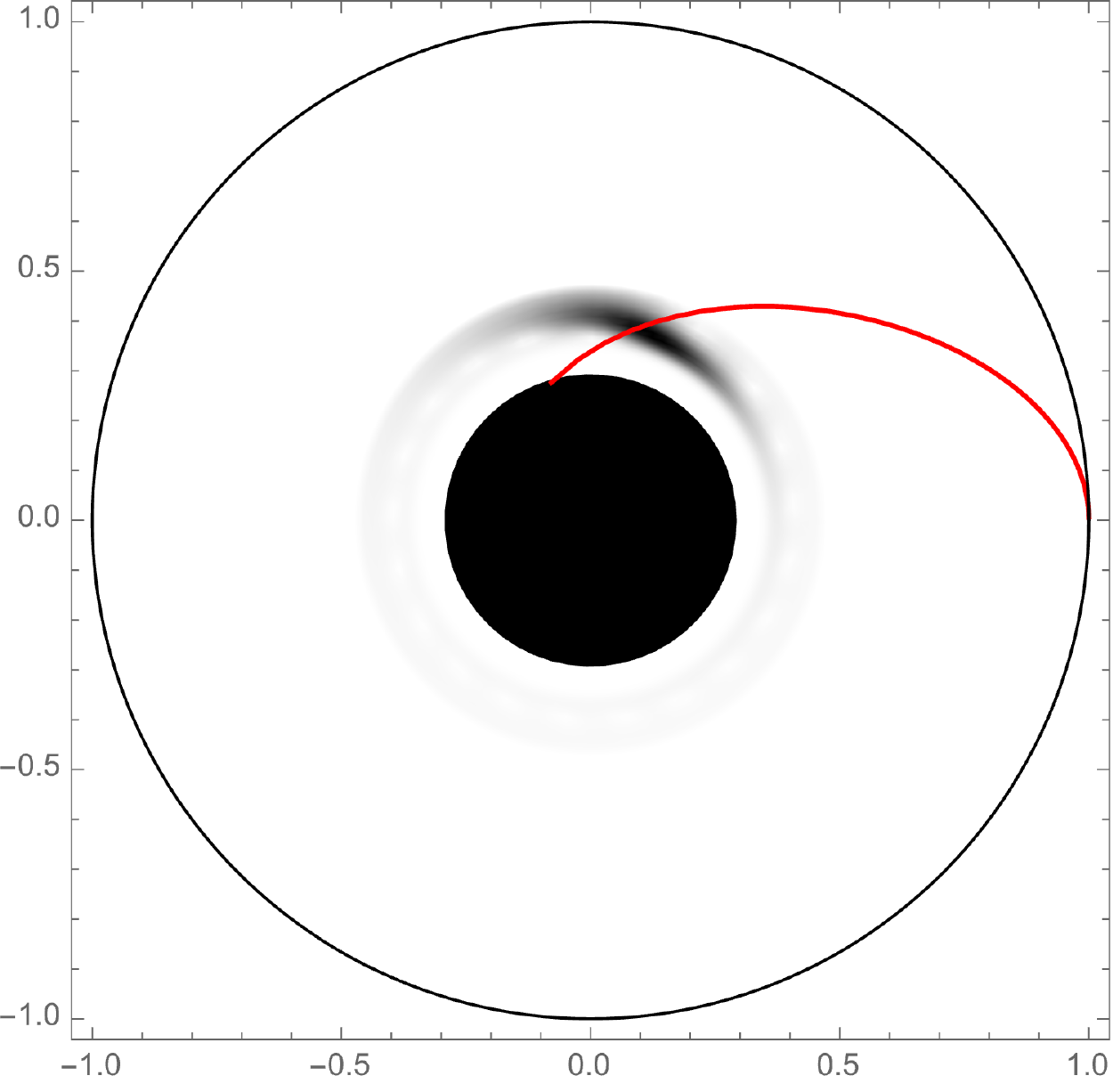}\\
        $t=3.15$
	\end{minipage}\\[24pt]
	\caption{Wave packets in AdS$_3$ and BTZ spacetime generated by the source~(\ref{eq: the source}). The horizontal and vertical axes are $x$ and $y$ defined in Eq.~(\ref{xy}).
    The darker the color is, the larger the absolute value of the scalar field is.
    The AdS radius is set $L=1$, and the other parameters are displayed above. The red curves are geodesic orbits specified by $m=M/\Omega$.
    }
	\label{fig: wave packet}
\end{figure}

Based on the idea, we numerically or analytically solve the equation of motion, $\nabla_\mu \nabla^\mu \Phi = 0$, not taking the eikonal approximation.
Hereafter, we will take the unit of $L=1$ in our actual calculations.
The numerical method is summarized in appendix~\ref{numerical}. 
In the case of AdS$_3$, we solved the equation analytically, and the process is shown in appendix \ref{app: AdS}.
The results of $d=2$ (pure AdS$_3$ and BTZ) are shown in Fig.~\ref{fig: wave packet}. The upper panels are for the pure AdS$_3$ and the lower panels are for BTZ with $r_h=0.3$. Parameters in the source are set as $M = 40$, $\Omega = 100$, $\sigma_t = 0.4$, $\sigma_\phi = 0.1$ for pure AdS$_3$ and 
$r_h=0.3$, $M = 30$, $\Omega = 50$, $\sigma_t = 0.3$, $\sigma_\phi = 0.1$ for BTZ.
Trajectories of null geodesics with specific angular-momentum $m = M/\Omega=0.4$ (pure AdS$_3$) and $m=0.3$ (BTZ) are shown by the red curves. We see that wave packets are generated by the boundary condition~(\ref{eq: the source}) and they move along the trajectories of null geodesics. For BTZ, the wave packet approaches the event horizon as we can expect from the analysis in section~\ref{sec: time lapse}: any null geodesics fall into the BTZ black hole. On the other hand, for AdS$_3$, the wave packet arrives at the antipodal point of the AdS boundary. When the wave packet arrives at the AdS boundary, we would get the pulse of the response function $\braket{O}_J$. 
This indicates that, under the Hawking-Page transition \cite{Hawking:1982dh}, which in $d=2$ is the transition between the BTZ and pure AdS, we observe the pulse of $\braket{O}_J$ only in the low temperature phase.

\begin{figure}[p]
    \centering
	Schwarzschild-AdS$_4$ ($r_h=0.3$, $m=0.83$, $M = 50$, $\Omega = 60$, $\sigma_t = 0.2, \sigma_\theta=\sigma_\phi = 0.1$)\\[12pt]
	\begin{minipage}{0.2\columnwidth}
         \centering
	    \includegraphics[width = 3.5cm]{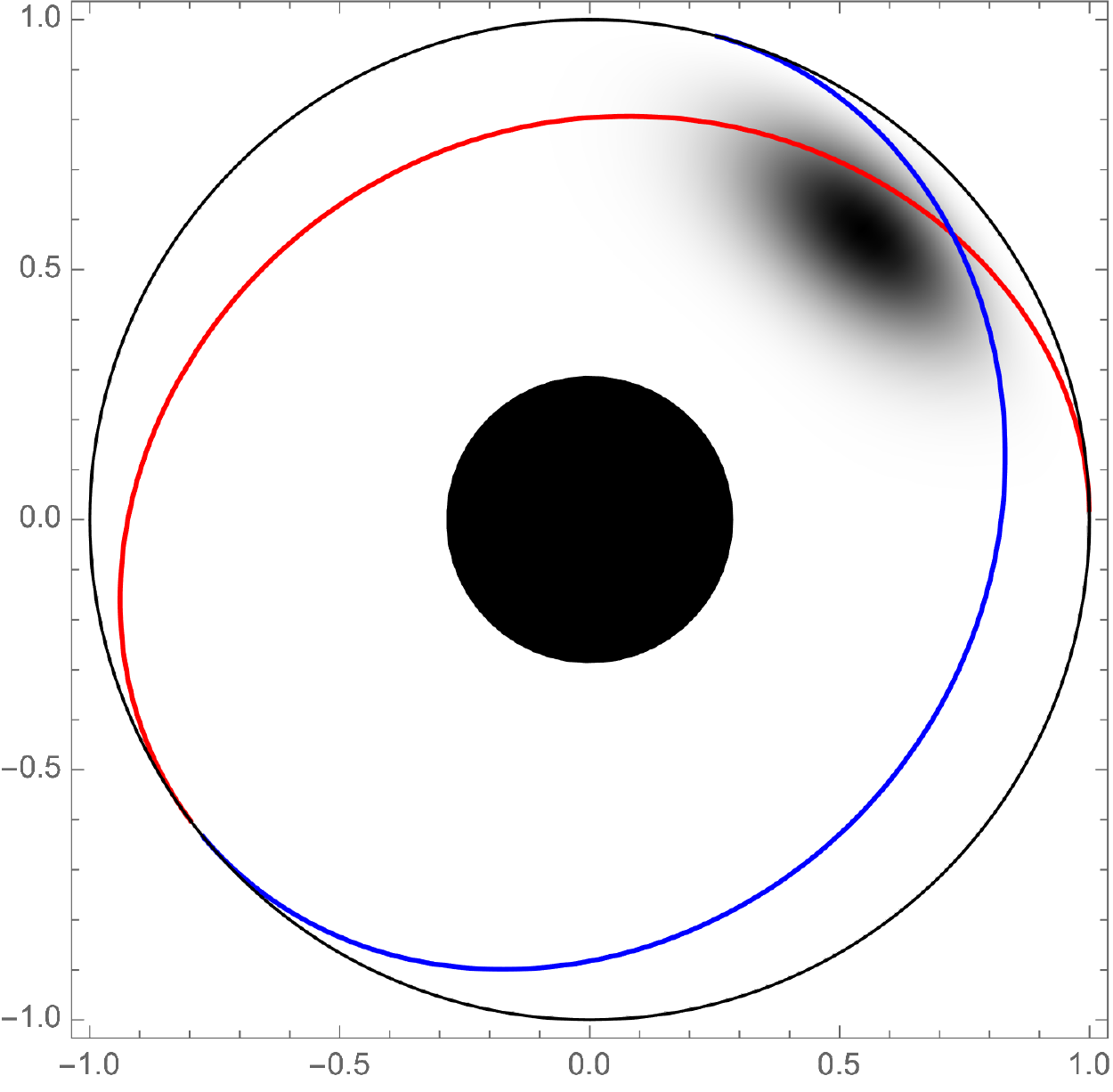}\\
        $t=1$
	\end{minipage}
     \hspace{6pt}
 	\begin{minipage}{0.2\columnwidth}
         \centering
	    \includegraphics[width = 3.5cm]{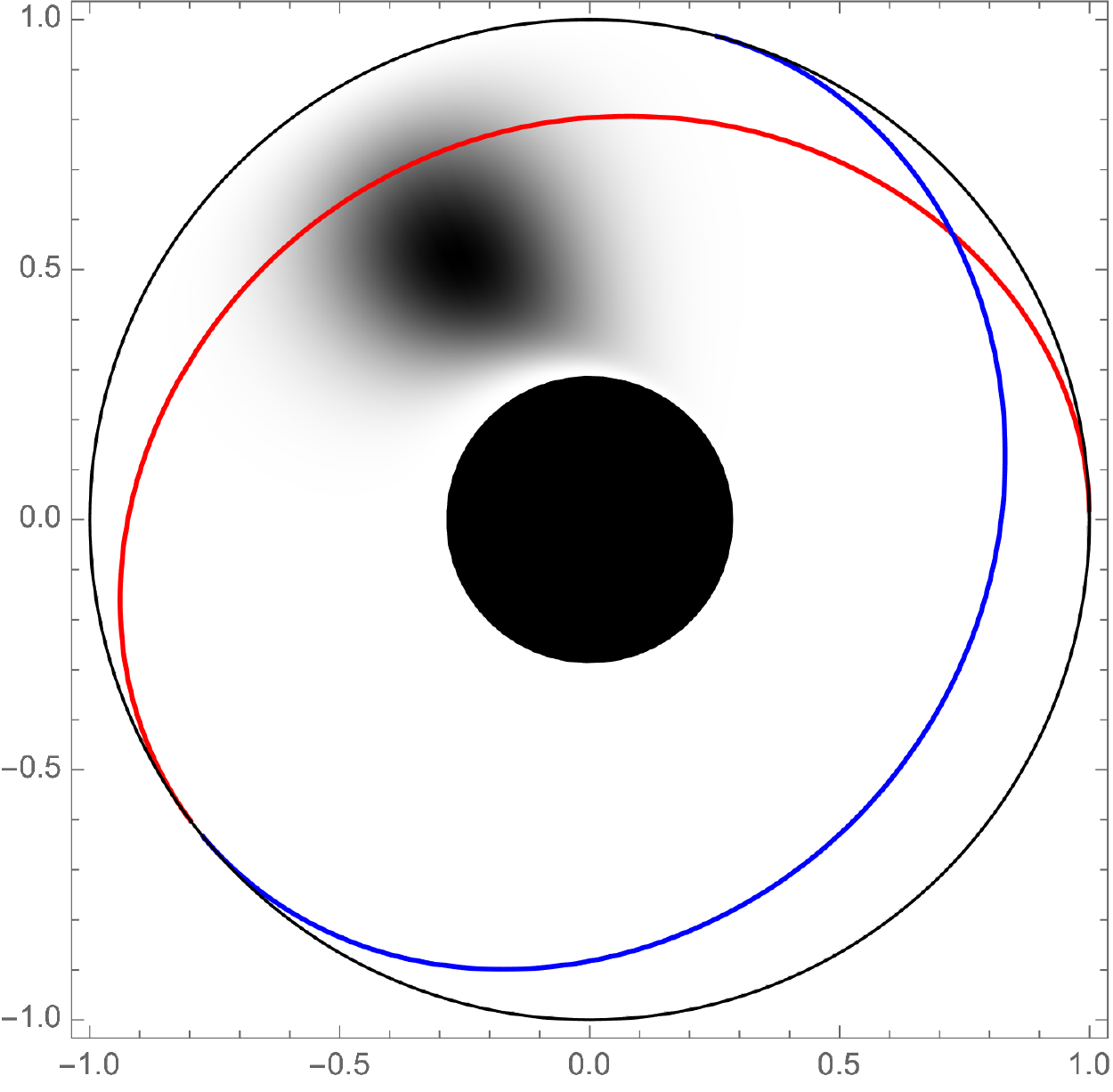}\\
        $t=2$
	\end{minipage}
     \hspace{6pt}
 	\begin{minipage}{0.2\columnwidth}
         \centering
	    \includegraphics[width = 3.5cm]{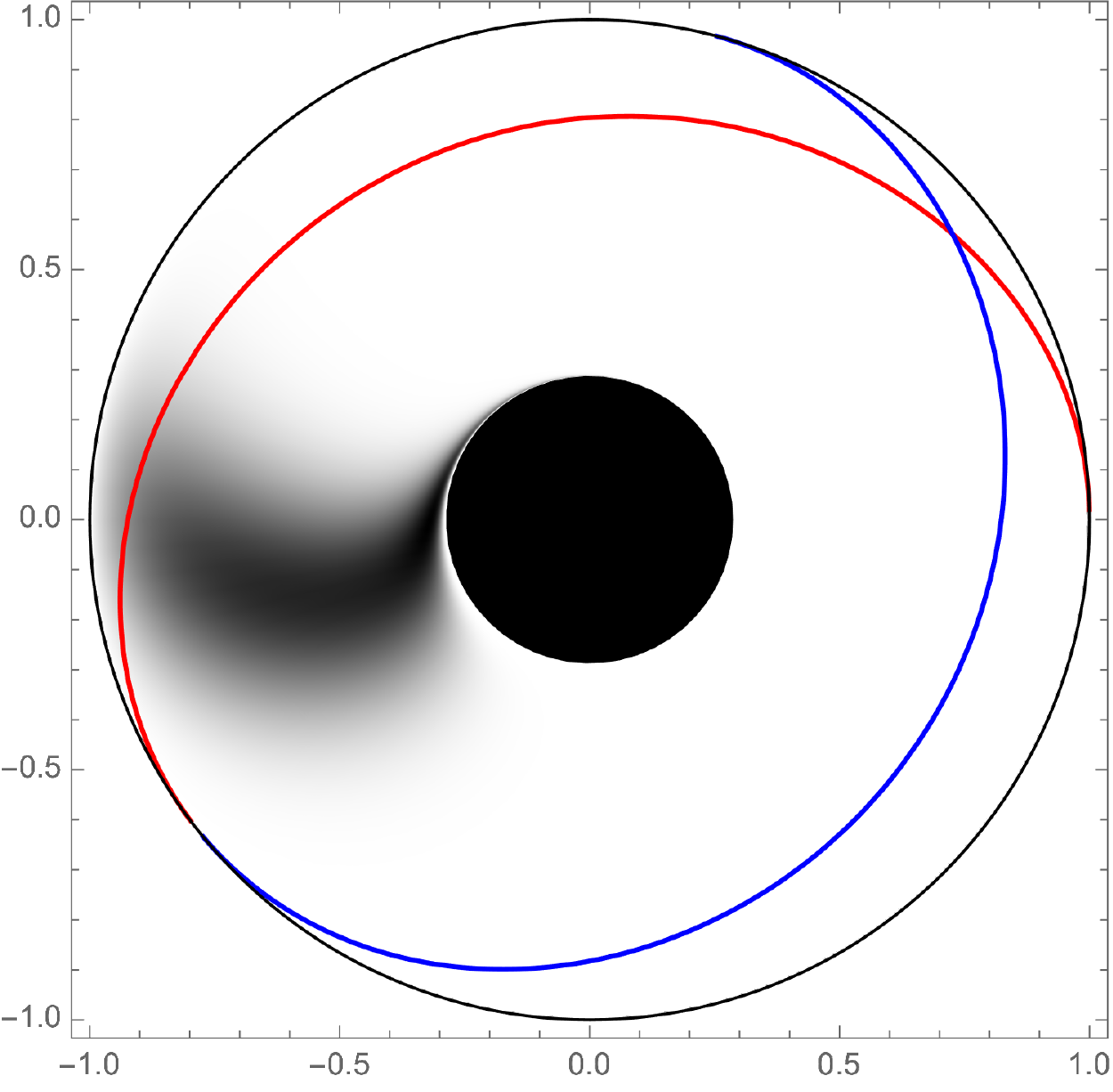}\\
        $t=3$
	\end{minipage}\\[24pt]
    
 	\begin{minipage}{0.2\columnwidth}
         \centering
	    \includegraphics[width = 3.5cm]{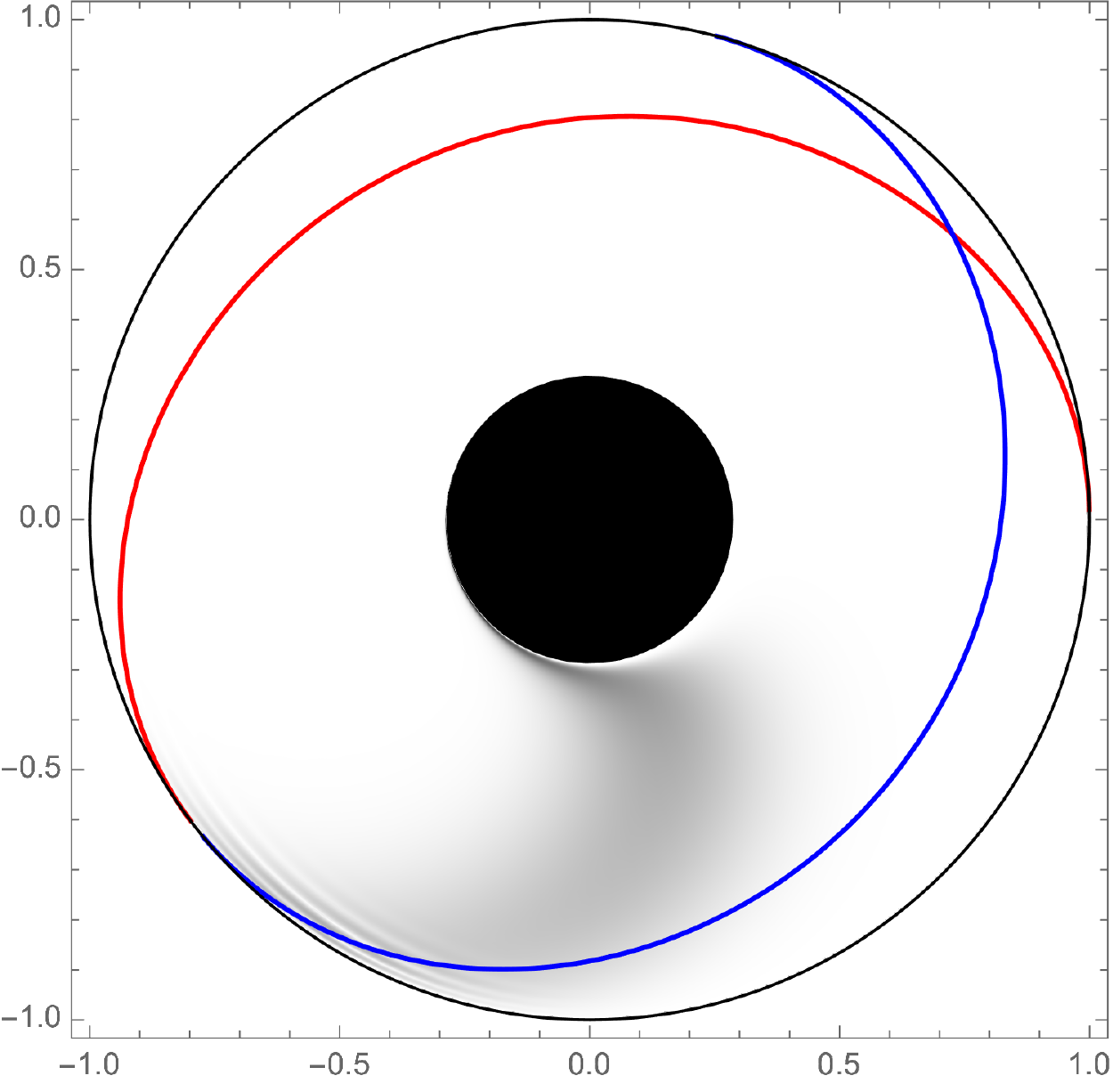}\\
        $t=4$
	\end{minipage}
\hspace{6pt}
	\begin{minipage}{0.2\columnwidth}
         \centering
	    \includegraphics[width = 3.5cm]{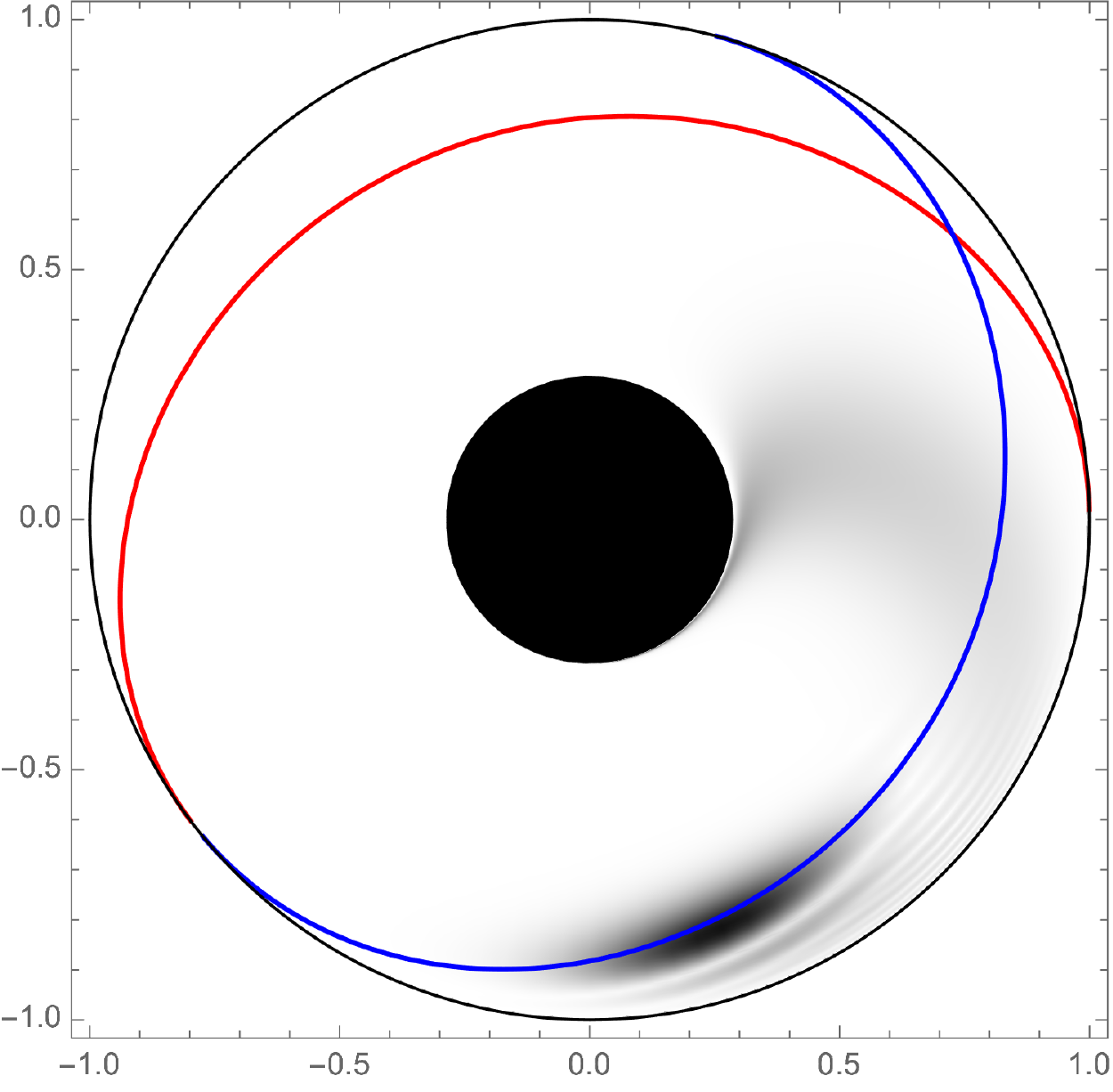}\\
        $t=5$
	\end{minipage}
     \hspace{6pt}
 	\begin{minipage}{0.2\columnwidth}
         \centering
	    \includegraphics[width = 3.5cm]{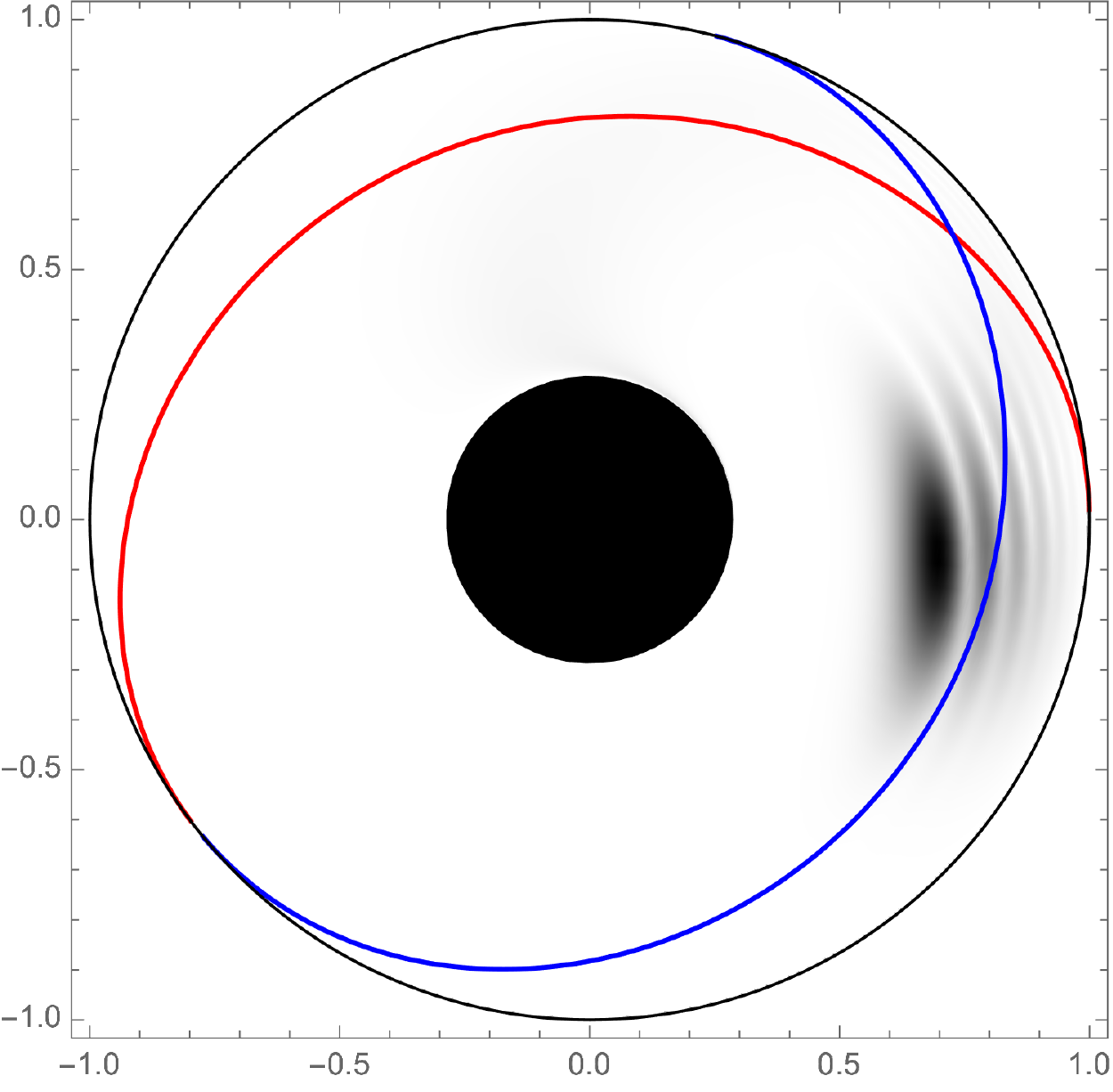}\\
         $t=6$
	\end{minipage}\\[24pt]
	\caption{The wave packet in Schwarzschild-AdS$_4$ generated by the source~(\ref{eq: the source4d}). Only equatorial plane $\theta=\pi/2$ is displayed. The horizontal and vertical axes are $x$ and $y$ defined in Eq.~(\ref{xy}).
    The darker the color is, the larger the absolute value of the scalar field is.
    The AdS radius is set $L=1$, and the other parameters are displayed above. The red curves are geodesic orbits specified by $m=M/\Omega$. The blue curves are represent geodesic orbits after the bounce at the AdS boundary. 
    }
	\label{fig: wave packet4d}
\end{figure}

The results of $d=3$ (Sch-AdS$_4$) are shown in Fig.~\ref{fig: wave packet4d}. 
Parameters are $r_h=0.3$, $M = 50$, $\Omega = 60$, $\sigma_t = 0.2$, and  $\sigma_\theta=\sigma_\phi = 0.1$. 
Trajectories of null geodesics with $m=M/\Omega=0.83$ are shown by the red and blue curves. The blue curves represent geodesics after the bounce at the AdS boundary. We see again that the wave packet moves along the trajectories of null geodesics. Unlike the BTZ, the wave packet can reach the AdS boundary depending on parameters. When the wave packet arrives at the AdS boundary, we would get the pulse of the response function $\braket{O}_J$ at $t=\Delta t$ and $\phi=\Delta \phi$. Our prediction of $(\Delta t,\Delta \phi)$ by using null geodesic is shown in Fig.~\ref{fig: plot of delta}.
If we observe the divergence of $\Delta t$ and $\Delta \phi$, it is an evidence of the existence of the photon sphere in the bulk.

% \begin{figure}
%     \centering
%     \includegraphics[width = 6cm]{AdS3_response.pdf}
%     \caption{The response function in the case of the AdS$_3$.
%     It suddenly stands up at around $t\sim \pi$ and $\phi \sim \pi$, as expected.
%     Note that the left and right vertical lines should be identified, due to the periodicity of the $\phi$-direction.}
%     \label{fig: response}
% \end{figure}

\begin{figure}
  \centering
\subfigure[AdS$_3$]
 {\includegraphics[scale=0.5]{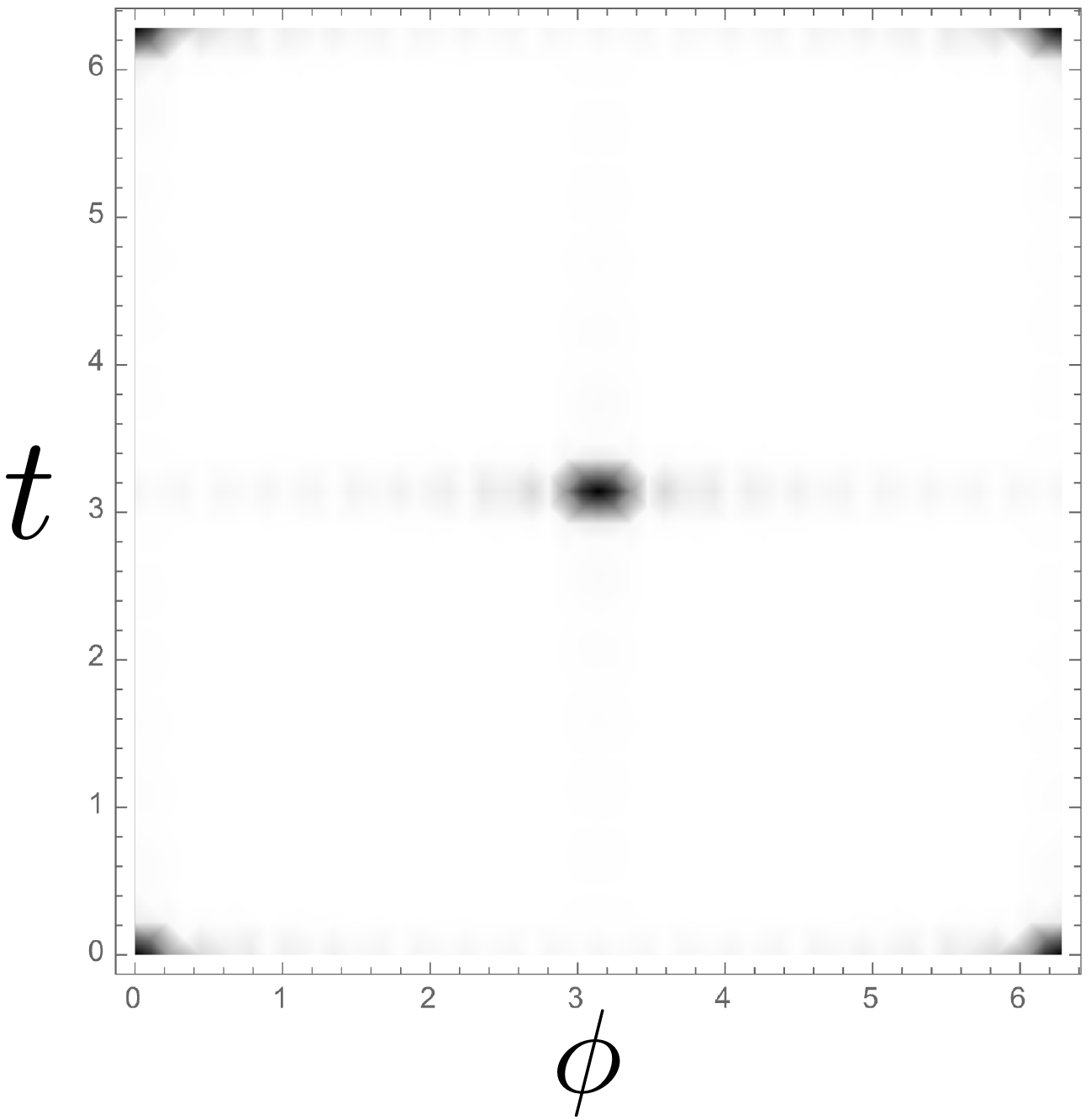}\label{fig: response}
  }
  \subfigure[Sch-AdS$_4$]
 {\includegraphics[scale=0.47]{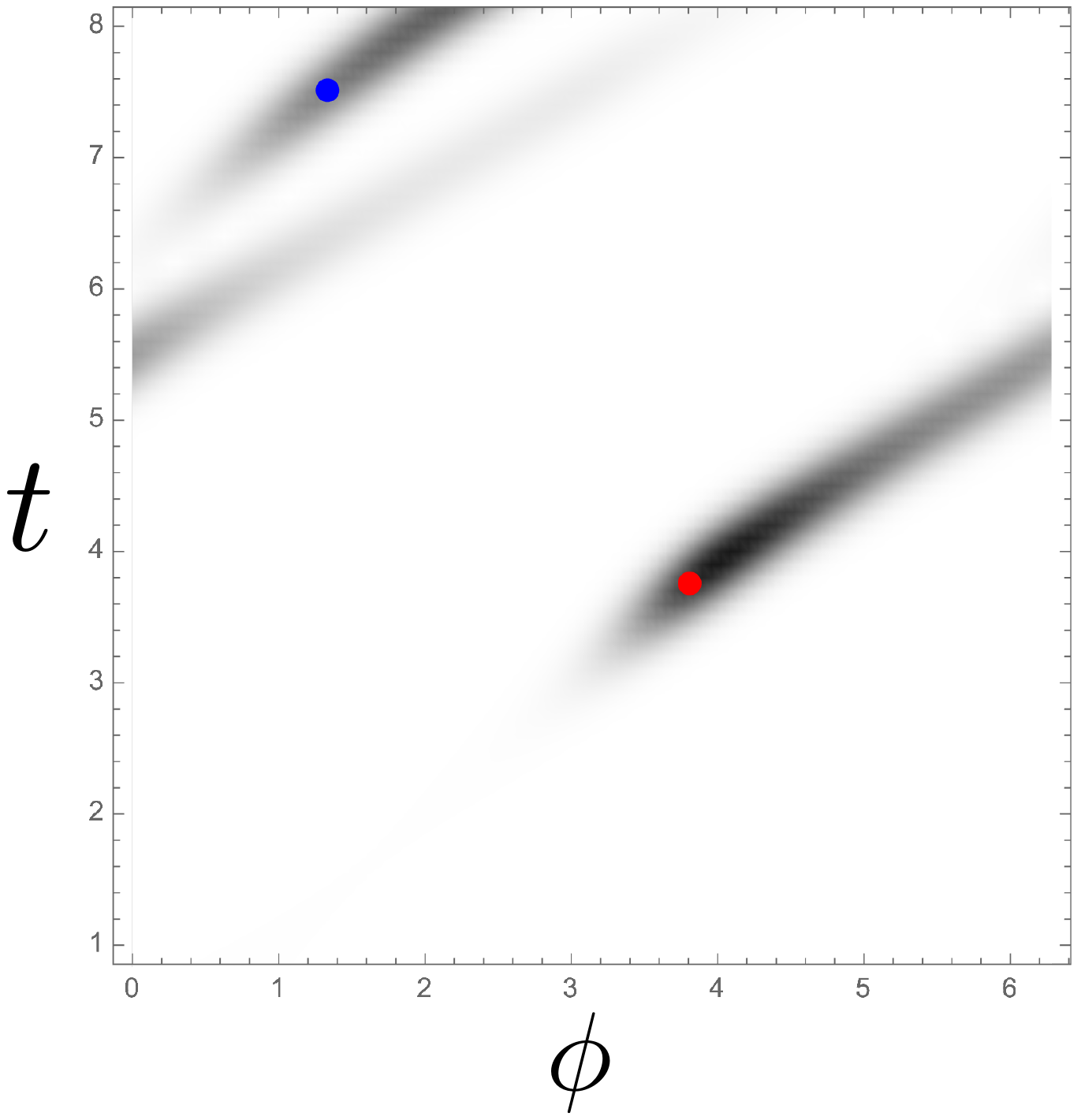}\label{fig: responseSch}
  }
 \caption{
    Response functions in the case of the (a) AdS$_3$ and (b) Sch-AdS$_4$.
    It suddenly stands up at around $t\sim \pi$ and $\phi \sim \pi$ for the AdS$_3$, as expected. For the Sch-AdS$_4$, the red and blue dots represent ($t,\phi$) at which the geodesic reached the boundary at the first and second times, respectively. 
    Note that the left and right vertical lines should be identified, due to the periodicity of the $\phi$-direction. Same parameters are used as in Figs. \ref{fig: wave packet} and \ref{fig: wave packet4d}.}
    \label{fig: responseAll}
\end{figure}

Finally, let us check the expectant behavior of the response function in the boundary theory.
Figure\ \ref{fig: response} shows response on the boundary $\braket{O}_J$ in the case of the AdS$_3$ 
when we provide the source $J$ given by Eq.~(\ref{eq: the source}) at around $(t,\phi)\sim (0,0)$. 
The computation is again shown in appendix \ref{app: AdS}. As we have expected, the response suddenly stands up just at the time the geodesic reaches the boundary, while it stays quiet at other times. We can also see that it again becomes large at $(t,\phi)\sim (2\pi,0)$, and this is because the wave packet comes back to $\phi = 0$, after bouncing at the antipodal point.
Figure\ \ref{fig: responseSch} shows the response in the case of the Sch-AdS$_4$. The source $J$ is given by Eq.~(\ref{eq: the source4d}) (see appendix~\ref{numerical}). We only display the response after applying the source, $t\geq 5\sigma_t$. The dots in the figure represent values of $(t,\phi)$ at which  the null geodesic arrives at the boundary: the red and blue dots correspond to the first and second arrivals. The response suddenly stands up at around the red dot as is the case with the AdS$_3$, while it has a longer tail comparing to the case of the AdS$_3$. This would be because of the diffusion of the wave packet of the bulk scalar field and a tidal disruption by the bulk black hole.
Since tidal disruptions are caused by the Weyl curvature of spacetimes in general, such a long tail of the response on the boundary might be a sign that a nontrivial dual geometry exists.

\section{Summary and discussions}\label{sec: summary}

We have devised a source to generate a wave packet propagating along a null geodesic inside the bulk, and verified that it works for several geometries, AdS$_3$, BTZ and Sch-AdS$_4$. In the gravity side, the wave packet propagates in the bulk for a while and then reaches the AdS boundary. When the wave packet arrive at the AdS boundary, we observe the pulse of the response $\braket{O}_J$. This is a natural behaviour of the null geodesic in AdS but gives a  peculiar prediction in the boundary theory: the response appears with a time lag. If there is a photon sphere in the bulk, the time lag can be infinite.
If this behaviour is detected against a quantum material in our world, that is a strong evidence of the material being holographic.

Collecting the data of $(\Delta t,\Delta\theta)$ for other typical geometries in the AdS/CFT would also be useful.
For example, it is known that holographic materials may exist among high-temperature superconductors.
Such materials, though their existence has not yet been confirmed, are called holographic superconductor.
In the dual spacetime of a holographic superconductor, a complex scalar field forms so-called ``scalar hair'' surrounding the AdS black hole, which will affect the orbits of null geodesics realized by the wave packets we have studied in section \ref{sec: shooting}.

The optical imaging to materials has been studied in \cite{Hashimoto:2018okj,Hashimoto:2019jmw}. 
They proposed that the holographic image of the AdS bulk can be constructed by the Fourier transformation of the response function with a window function on the boundary theory.
The optical imaging for the null geodesic created in this paper is an interesting future direction. By the imaging, we would be able to determine the incident angle of the null geodesic to the AdS boundary.
Whereas the broad sources containing various angular momenta were used in \cite{Hashimoto:2018okj,Hashimoto:2019jmw}, our source has been highly localized even in the momentum space.
Thus, we would obtain a spot-like image deflected by gravitational potential in the bulk.

In the cases of the holographic material, when we provide the source proposed in this paper, we can ideally observe no response of the corresponding one-point function while the null geodesic is wandering inside the bulk.
However, since the total energy of the system should be conserved, the material has been excited in spite of no response of the one-point function.
Observing multi-point functions (two-point, three-point, and so on) on the boundary may allow us to probe null geodesics deep inside the bulk.

%エネルギーは高くなってる。でも応答はない。不思議。
%でも、多点関数をみれば、bulkを漂う測地線の存在を見ることができるだろう。これはfuture work.

\subsection*{Acknowledgement}
We thank Koji Hashimoto and Takuya Yoda for discussions.
The work of S.K. was supported in part by JSPS KAKENHI Grant No. 16K17704.
The work of K.M. was supported in part by JSPS KAKENHI Grant Nos. 20K03976, 21H05186 and 22H01217.
The work of D.T.\ is supported by Grant-in-Aid for JSPS Fellows No.\ 22J20722.

%%%%%
\appendix

\section{Numerical detail}
\label{numerical}

\begin{figure}[t]
\centering
\includegraphics[scale=0.6]{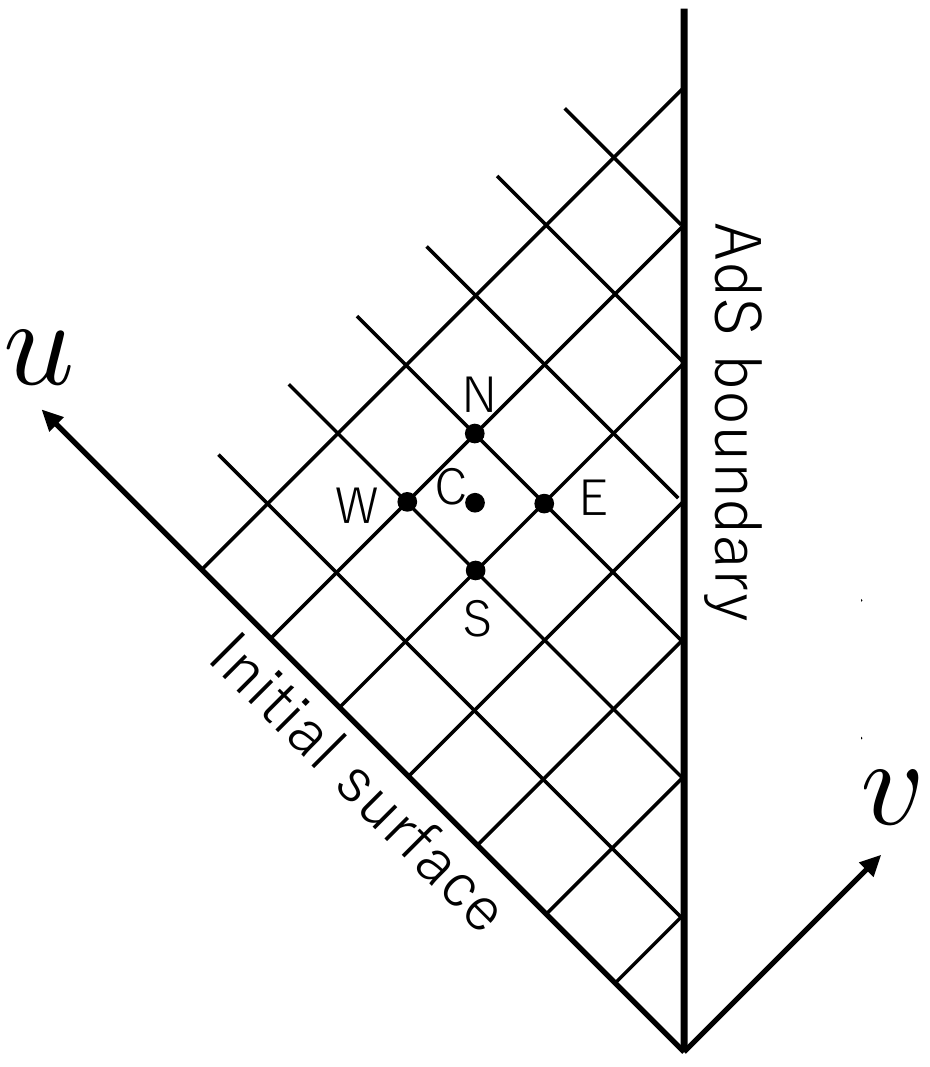}
\caption{Discretization of the 2-dimensional spacetime spanned by $(u,v)$-coordinates.
}
\label{fig:disc}
\end{figure}

We explain the numerical method to solve the time evolution of the scalar field in asymptotically AdS spacetimes~(\ref{eq: metric}).
Here, we focus only on the scalar field in Sch-AdS$_4$ for concreteness. We can easily apply the same technique to the BTZ spacetime.
Decomposing the scalar field as $\Phi=\sum_{l,m} c_{lm}\varphi_l(t,x) Y_{lm}(\theta,\phi)$ by the spherical harmonics, we have the wave equation in ($1+1$) dimensions as
\begin{equation}
 \left[
-\partial_t^2+\partial_x^2 + \frac{2f(r)}{r}\partial_x -\frac{l(l+1)f(r)}{r^2}
\right]\varphi_l(t,x)=0\ ,
\label{2deq}
\end{equation}
where we have introduced the tortoise coordinate 
\begin{equation}
 x=\int_{\infty}^r \frac{dr'}{f(r')}\ .
\end{equation}
In this coordinate the AdS boundary and the event horizon are located at $x=0$ and $x=-\infty$, respectively. We further introduce double null coordinates as
\begin{equation}
 u=\frac{(t-t_0)-x}{2}\ ,\quad v=\frac{(t-t_0)+x}{2}\ ,
\end{equation}
where $t_0(<0)$ denotes an initial time.
Note that the origin of ($u,v$) lies on $t=t_0$ and $x=0$.
Then, Eq.~(\ref{2deq}) is written as
\begin{equation}
 \left[
-\partial_u \partial_v + \frac{f(r)}{r}(\partial_v-\partial_u) -\frac{l(l+1)f(r)}{r^2}
\right]\varphi_l(u,v)=0\ .
\label{uveq}
\end{equation}
We discretize coordinates $(u,v)$ as in Fig.~\ref{fig:disc}.
For instance, let us focus on points N, E, W, S, and C in the figure. The scalar field $\varphi_l$ and its derivatives at the point C are written as\footnote{
We found that the other discretization  $\varphi_l = (\varphi_l^E+\varphi_l^W)/2$ causes the numerical instability for the scalar field in the Sch-AdS$_4$. For the BTZ spacetime, both choice was numerically stable. 
}
\begin{equation}
\begin{split}
 &\partial_u \partial_v \varphi_l = \frac{\varphi_l^N-\varphi_l^E-\varphi_l^W+\varphi_l^S}{h^2}\ ,\quad
 \partial_u \varphi_l = \frac{\varphi_l^N-\varphi_l^E+\varphi_l^W-\varphi_l^S}{2h}\ ,\\
 &\partial_v \varphi_l = \frac{\varphi_l^N+\varphi_l^E-\varphi_l^W-\varphi_l^S}{2h}\ ,\quad 
 \varphi_l = \frac{\varphi_l^N+\varphi_l^S}{2}\ .
\end{split}
\end{equation}
where $\varphi_l^{N,E,W,S}$ are values of the scalar field at N, E, W, and S and $h$ is the step size. The above discretization is the second order accuracy in $h$.
Substituting the above expressions into Eq.~(\ref{uveq}), we have the equation to determine $\varphi_l^N$ from $\varphi_l^{E,W,S}$.
Thus, once we give data of the scalar field at the initial surface ($v=0$) and the AdS boundary ($u=v$), we can determine the dynamics of the scalar field in their domain of dependence. At the initial surface, we set $\varphi_l|_{v=0}=0$. At the AdS boundary, we impose
\begin{equation}
 \varphi_l|_{u=v}=\frac{1}{(2\pi)^{1/2} \sigma_t}\exp\left[-i\Omega t -\frac{t^2}{2\sigma_t^2}\right] w(t)\ ,
\end{equation}
where we assume $\sigma_t \ll |t_0|$ to be $\varphi_l|_{u=v} \simeq 0$ at the initial time $t=t_0$. In our actual numerical calculation, we have introduced the window function $w(t)$ defined by 
\begin{equation}
    w(t)=
    \begin{cases}
    \sin^2\left(\frac{\pi (t-t_1)}{t_2-t_1}\right) & (t_1 < t < t_2) \\
    0 & (\textrm{otherwise})
    \end{cases}\ ,
\end{equation}
in order that the source function has compact support.
We set $t_1=-5\sigma_t$ and $t_2=5\sigma_t$.
From numerical solutions of $\varphi_l$, we obtain the solution in the position space as
\begin{equation}
 \Phi(t,r,\theta,\phi)=\sum_{l=0}^\infty \sum_{m=-l}^l c_{lm}\,  \varphi_l(t,x) Y_{lm}(\theta,\phi)\ ,
\end{equation}
where $c_{lm}$ is a constant. To realize the boundary condition~(\ref{eq: the source4d}), we choose the constant coefficient $c_{lm}$ as
\begin{equation}
\begin{split}
 c_{lm}&=\frac{1}{2\pi \sigma_\theta \sigma_\phi}\int_0^{2\pi}d\phi \int_{-1}^1 d\cos\theta \exp\left[i M \phi-\frac{(\theta-\pi/2)^2}{2\sigma_\theta^2}-\frac{\phi^2}{2\sigma_\phi^2}\right] Y_{lm}^\ast (\theta,\phi)\\
&\simeq \sqrt{\frac{2l+1}{4\pi} \frac{(l-|m|)!}{(l+|m|)!}}\frac{(l+|m|-1)!!}{(l-|m|)!!}  \\
&\hspace{3cm}\times \exp\left[
-\frac{\sigma_\theta ^2}{2}\left(l(l+1)-m^2+\frac{1}{2}\right)-\frac{\sigma_\phi^2}{2}(m-M)^2
\right]\ .
\end{split}
\end{equation}
At the second equality, we have used $\sigma_\theta, \sigma_\phi \ll 1$.

After applying the source $(t>t_2)$, the numerical solution can be expanded as
\begin{equation}
    \varphi = p_3(v) (u-v)^3 + p_4(v) (u-v)^4 + \cdots\ .
\end{equation}
Fitting the numerical solution by the fourth order polynomial in $u-v$ near the AdS boundary, we obtain $p_3(v)$. The response is then computed as 
$\langle \mathcal{O} \rangle_J = -p_3(v)$.

\section{Analytic computation in AdS\texorpdfstring{$_3$}{TEXT}}\label{app: AdS}
Here, on the background of AdS$_3$, we analytically solve the equation of motion of \eqref{eq: action},
\begin{align}
	\frac{1}{\sqrt{-g}}\partial_\mu (\sqrt{-g}g^{\mu \nu }\partial_\nu \Phi) = 0,\label{eq: EOM}
\end{align}
with the boundary condition based on the GKPW dictionary
\begin{align}
	\Phi(t,\infty,\phi) = J(t,\phi),\label{eq: BC}
\end{align}
where $J$ is given in \eqref{eq: the source}.
%Equation \eqref{eq: BC} is based on the GKPW dictionary.
We set $L=1$ in this appendix.

We first write $\Phi$ as
\begin{align}
	\Phi(t,r,\phi) = \frac{1}{(2\pi)^2}\sum_{n=-\infty}^{\infty}\int \d \omega\,e^{-i\omega t+i n\phi} f_{\omega,n}(r),
\end{align}
and plug this into \eqref{eq: EOM}:
\begin{align}
	\frac{d^2}{dr^2}f_{\omega,n}(r) + \frac{3r^2+1}{r^2+1} \frac{d}{dr}f_{\omega,n}(r)+\l[\frac{\omega^2}{(r^2+1)^2}-\frac{n^2}{r^2(r^2+1)}\r]f_{\omega,n}(r) = 0.
	\label{eq: EOM for f}
\end{align}
Two independent solutions for \eqref{eq: EOM for f} are given as
\begin{align}
	\xi^{n}F\left(\frac{\omega+n}{2},-\frac{\omega-n}{2},1+n,\xi^2 \right)
	,\qquad
	\xi^{-n}F\left(\frac{\omega-n}{2},-\frac{\omega+n}{2},1-n,\xi^2 \right),
\end{align}
where $F$ is the hypergeometric function and $\xi := r/\sqrt{r^2+1}$.
Since the wave should not diverge at $r=0$ (or $\xi = 0$), we adopt the following as $f_{\omega,n}$:
\begin{align}
	f_{\omega,n}(r) = C_{\omega,n}\xi^{|n|}F\left(\frac{\omega+|n|}{2},-\frac{\omega-|n|}{2},1+|n|,\xi^2 \right).
\end{align}
Here $C_{\omega,n}$ is a constant which may depend on $\omega$ and $n$.

The solution we have now is
\begin{align}
	\Phi(t,r,\phi) = \frac{1}{(2\pi)^2}\sum_{n=-\infty}^{\infty}\int \d \omega\,e^{-i\omega t+i n\phi}
	C_{\omega,n}\xi^{|n|}F\left(\frac{\omega+|n|}{2},-\frac{\omega-|n|}{2},1+|n|,\xi^2 \right).
\end{align}
To determine $C_{\omega,n}$, we use \eqref{eq: BC}, which turns out to be
\begin{align}
	\frac{1}{(2\pi)^2}\sum_{n=-\infty}^{\infty}\int \d \omega\,
	\frac{C_{\omega,n}\Gamma(|n|+1)e^{-i\omega t+i n\phi}}{\Gamma(1+\frac{|n|-\omega}{2})\Gamma(1+\frac{|n|+\omega}{2})}
	=
	 \frac{1}{2\pi \sigma_t \sigma_\phi}\exp\left[-i\Omega t + i M \phi
	-\frac{t^2}{2\sigma_t^2}-\frac{\phi^2}{2\sigma_\phi^2}
	 \right].
\end{align}
By Fourier-transforming this with $\sigma_t, \sigma_\phi \ll 1$, we have
\begin{align}
	C_{\omega,n} = \frac{\Gamma(1+\frac{|n|-\omega}{2})\Gamma(1+\frac{|n|+\omega}{2})}{\Gamma(|n|+1)}
	\exp\left[-\frac{\sigma_t^2}{2}(\omega-\Omega)^2-\frac{\sigma_\phi^2}{2}(n-M)^2 \right],
\end{align}
and hence the solution is finally
\begin{align}
	\Phi(t,r,\phi) = \frac{1}{(2\pi)^2}&\sum_{n=-\infty}^{\infty}\int \d \omega\,e^{-i\omega t+i n\phi}
	\xi^{|n|}F\left(\frac{\omega+|n|}{2},-\frac{\omega-|n|}{2},1+|n|,\xi^2 \right)\nonumber\\
	&\times\frac{\Gamma(1+\frac{|n|-\omega}{2})\Gamma(1+\frac{|n|+\omega}{2})}{\Gamma(|n|+1)}
	\exp\left[-\frac{\sigma_t^2}{2}(\omega-\Omega)^2-\frac{\sigma_\phi^2}{2}(n-M)^2 \right].\label{eq: AdS solution}
\end{align}

In the solution, there are first order poles along the real axis of $\omega$, at $\omega = \pm(2k+|n|)$ ($k = 1,2,\cdots$).
To perform the integration,  we slightly move them down to the imaginary direction, adding $-i 0$.
This corresponds to adopting a boundary condition that $\phi$ vanishes in the past, $t<0$.
Therefore, from the residue theorem, we obtain
\begin{align}
	\Phi(t,r,\phi) = -2\pi i& \sum_{n=-\infty}^{\infty}\sum_{k=1}^{\infty}\sum_{s=\pm1}
	\exp\left[
	-\frac{\sigma_t^2}{2}\Bigl((2k+|n|)s-\Omega\Bigr)^2-\frac{\sigma_\phi^2}{2}(n-M)^2
	 \right]\nonumber\\
	 &\times e^{-is(|n|+2k)t+in\phi}(-1)^k\frac{(|n|+k+1)!}{|n|!k!}\xi^{|n|}F(|n|+k,-k,1+|n|,\xi^2).
\end{align}

Let us read the response function on the boundary theory.
The asymptotic form of the above solution is,
\begin{align}
    \Phi(t,r,\phi) = -2\pi i& \sum_{n=-\infty}^{\infty}\sum_{k=1}^{\infty}\sum_{s=\pm1}
	\exp\left[
	-\frac{\sigma_t^2}{2}\Bigl((2k+|n|)s-\Omega\Bigr)^2-\frac{\sigma_\phi^2}{2}(n-M)^2
	 \right]\nonumber\\
	 &\times e^{-is(|n|+2k)t+in\phi}(|n| + k + 1)(|n|+k) (1-\xi)+\cdots.\label{eq: response}
\end{align}
Usually in the AdS$_3$, $O(1)$ or $O(r^{-2}\ln r)$ terms from non-normalizable modes appear in the asymptotic expression, but this time no such terms appear.
This is because picking up poles in \eqref{eq: AdS solution} is equivalent to expanding the solution in terms of normalizable modes, which physically means that the source on the boundary is soon turned off and only normalizable modes remain excited inside the bulk.
Therefore, recalling what the GKPW dictionary says, we regard the coefficient of $(1-\xi)$ in \eqref{eq: response} as the response function.

%%%%%TODO bibliography
\bibliography{ref}

%似たセンス
%\cite{Dempsey:2016wad}
%\bibitem{Dempsey:2016wad}
%D.~Dempsey and S.~R.~Dolan,
%``Waves and null congruences in a draining bathtub,''
%Int. J. Mod. Phys. D \textbf{25}, no.09, 1641004 (2016)
%doi:10.1142/S0218271816410042
%[arXiv:1602.07992 [gr-qc]].

\end{document}